\documentclass[pra,superscriptaddress,preprintnumbers,preprint,showpacs,amsmath,amssymb]{revtex4-1}
\usepackage{morefloats}
\usepackage{psfrag}
\usepackage{graphicx}
\usepackage{subfigure}
\usepackage{array}
\usepackage{natbib} 
\def\be{\begin{equation}}       \def\ee{\end{equation}}
\def\bea{\begin{eqnarray}}      \def\eea{\end{eqnarray}}

\begin{document}
\title{Manipulating entanglement sudden death in two coupled two-level atoms interacting off-resonance with a radiation field: an exact treatment}
\author{Gehad Sadiek\footnote{Corresponding author: gsadiek@sharjah.ac.ae}}
\affiliation{Department of Applied Physics and Astronomy, University of Sharjah, Sharjah 27272, UAE}
\affiliation{Department of Physics, Ain Shams University, Cairo 11566, Egypt}
\author{Wiam Al-Drees}
\affiliation{Department of Physics and Astronomy, King Saud University, Riyadh 11451, Saudi Arabia}
\author{M. Sebaweh Abdallah\footnote{Deceased.}}
\affiliation{Department of Mathematics, King Saud University, Riyadh 11451, Saudi Arabia}
\begin{abstract}
We study a model of two coupled two-level atoms (qubits) interacting off-resonance (at non-zero detuning) with a single mode radiation field. This system is of special interest in the field of quantum information processing (QIP) and can be realized in electron spin states in quantum dots or Rydberg atoms in optical cavities and superconducting qubits in linear resonators. We present an exact analytical solution for the time evolution of the system starting from any initial state. Utilizing this solution, we show how the entanglement sudden death (ESD), which represents a major threat to QIP, can be efficiently controlled by tuning atom-atom coupling and non-zero detuning.  We demonstrate that while one of these two system parameters may not separately affect the ESD, combining the two can be very effective, as in the case of an initial correlated Bell state. However in other cases, such as a W-like initial state, they may have a competing impacts on ESD. Moreover, their combined effect can be used to create ESD in the system as in the case of an anti-correlated initial Bell state. A clear synchronization between the population inversion collapse-revival pattern and the entanglement dynamics is observed at all system parameter combinations. Nevertheless, only for initial states that may evolve to ESD, the population inversion revival oscillations, where exchange of energy between the atoms and the field takes place, temporally coincide with the entanglement revival peaks, whereas the population collapse periods match the ESD intervals.  The variation of the radiation field intensity has a clear impact on the duration of the ESD at any combination of the other system parameters.
\end{abstract}
\maketitle
\section{Introduction}
\label{Introduction}
The great interest in realizing quantum information processing (QIP) systems in the last few decades, capable of performing efficient quantum simulation and quantum computing tasks \cite{Nielsen-Chuang2010}, led to a significant progress in engineering new quantum systems that are considered very promising candidates for playing the role of a qubit. These developed artificial atomic systems (such as semiconducting quantum dots and superconducting circuits) in addition to customized natural atomic systems (such as Rydberg atoms and trapped atoms, ions and molecules), in contrary to the natural conventional atoms, enjoy a strong coupling with a similar type of system or even with a different type (when implemented in a hybrid system) \cite{Buluta:2011, Xiang:2013, Lodahl:2015, Wendin:2017}, through direct or mediated interaction. A significant breakthrough in the role of artificial atomic systems in the QIP arena was achieved when superconducting qubits were successfully embedded in a superconducting microwave resonator \cite {Yang:2003,Yang:2004,You:2003,Kis:2004}, paving the way for the new paradigm of circuit quantum electrodynamics (cQED). Another important step came out when an architecture for quantum computation using cQED was introduced by Balis {\it et al} \cite{Blais:2004}, where they proposed to use a one-dimensional (1D) transmission line resonator consisting of a full-wave section of superconducting coplanar wave guide to play the role of a cavity and a superconducting qubit as the atom. They showed that this structure can be customized to access the strong coupling regime between the field and the qubit in analogy to what can be achieved in optical cavity quantum electrodynamics (CQED). Furthermore, they demonstrated that the proposed architecture can be efficiently utilized for the coherent control, entanglement, and readout of qubits in QIP. Particularly, they emphasized the possibility of generating tunable entanglement between two qubits that are few centimeters apart within the resonator, where they introduced, in addition to the qubits-field interaction, an effective coupling between the two qubits mediated by the virtual excitation of the resonator. Soon after, a strong coupling between a single photon and a superconducting qubit using cQED was realized experimentally \cite{Wallraff:2004}. Several experimental works demonstrated that two close superconducting qubits can be directly coupled via local interactions \cite{Yamamoto:2003, Berkley:2003, Majer:2005, Steffen:2006, Hime:2006, Van:2007, Niskanen:2007}. Latter, the coupling of two distant superconducting qubits mediated by microwave photons confined in a superconducting transmission line was reported \cite{Majer:2007}. In fact, proposals for similar schemes in cavity QED was introduced and implemented particularly for coupling (directly or indirectly) Rydberg atoms inside optical cavities \cite{Hagley:1997,Zheng:2000,Raimond:2001,Osnaghi:2001,Gywat:2006,Saffman:2010,Guerlin:2010,Donaire:2017}.

Currently, there is a great progress in developing superconducting multi-qubit circuits with long coherence time embedded in 1D, 2D and 3D superconducting resonators, which can perform high-fidelity quantum gates; extended review and references can be found in \cite{Wendin:2017}. Remarkably, not only superconducting qubits can be integrated into superconducting resonators but also atoms (trapped and Rydberg) and spins (in Quantum dots or solid state impurities) producing hybrid circuit QED \cite{Xiang:2013, Wendin:2017}. In these hybrid circuits, superconducting qubits can have either direct interaction with atoms (or spins) through electric or magnetic fields or indirect one mediated by the radiation field. These hybrid structures are of great interest for QIP as it combines the advantages of the two sides, insensitivity and long coherence time of atoms (spins) and rapid processing in superconducting circuits. Very recently, Nguyen {\it et al} \cite{Nguyen:2018} have proposed a new approach for analogue quantum simulation of spin arrays based on laser-trapped circular Rydberg atoms, benefiting from their long coherence time and insensitivity to collisions and photo-ionization, relying on the available state-of-the-art experimental techniques. They showed that the strong coupling between the atomic dipoles can be utilized to simulate a spin-1/2 $XXZ$ chain Hamiltonian with fully tunable nearest neighbor coupling over a wide range and studied the adiabatic time evolution of the chain. Furthermore, they suggested that this scheme can be implemented in CQED to overcome many of the challenges in Rydberg-atom CQED or even can be utilized in Hybrid cQED experiments by integrating superconducting circuits and laser-trapped Rydberg atoms.

In general, for a composite system with several interacting qubits (of same type or different) coupled to a single radiation mode, the Hamiltonian assumes the form \cite{Blais:2004,Gywat:2006,Xiang:2013, Wendin:2017} 
\be
H = \Omega \; \hat{a}^\dagger \hat{a} + \frac{1}{2} \sum_{i} \; \omega_{i} \; \hat{\sigma}^{(i)}_{z}\; 
+  \sum_{i} \lambda_{1}^i \; \hat{\sigma}^{(i)}_{x}(\hat{a}+\hat{a}^\dagger) \; 
+ \sum_{ij,\nu} \lambda_{2,\nu}^{ij} \; \hat{\sigma}^{(i)}_{\nu}\hat{\sigma}^{(j)}_{\nu} \;
\label{H_multi-qubit}
\ee
The first and second terms in the Hamiltonian represent the free quantized radiation field and the non-interacting qubits while the third and fourth terms represent the qubit-field and qubit-qubit interactions respectively. $\Omega$ and $\omega_{i}$ are the frequencies of the single-mode radiation field and the $i$th qubit transition respectively, $\hat{a}^\dagger$ and $\hat{a}$ are creation and annihilation operators of the radiation field which satisfy the usual commutation relation $[\hat{a},\hat{a}^\dagger]=1$ and $\hat{\sigma}^{(i)}_{\nu}$, where $\nu=x,y,z$, are the usual Pauli spin operators representing the $i$th qubit. There are different regimes of the coupling strength between the radiation field and the qubit, which are mainly determined by the ratio between $\lambda_{1}^i$ and the other energy scales of the composite system namely $\Omega$, $\omega_i$, $\kappa$ and $\gamma_i$, where $\kappa$ and $\gamma_i$ are the decay rates of the resonator (cavity) and the $i$th qubit respectively. When $\lambda_{1}^i << \Omega, \omega_i, \kappa,\gamma_i$, the system is in the weak coupling regime, whereas for $\kappa,\gamma_i << \lambda_{1}^i << \Omega, \omega_i\;$, it is in the strong coupling regime. In these regimes the rotated wave approximation (RWA)~\cite{Scully-Zubairy1997, Irish:2007} is valid, which when applied converts the third term in the Hamiltonian into the usual Jaynes-Cummings form $\sum_{i} \lambda_{1}^i \; (\hat{a}\hat{\sigma}^{(i)}_{+}+\hat{a}^\dagger \hat{\sigma}^{(i)}_{-})$, where $\hat{\sigma}^{(i)}_{\pm}=\hat{\sigma}^{(i)}_{x} \pm \hat{\sigma}^{(i)}_{y}$. The RWA fails in the ultra-strong and deep-strong regimes where $\lambda_{1}^i$ becomes of the same order of magnitude as $\Omega$ or higher. The qubit-qubit coupling constant $\lambda_{2,\nu}^{ij}$ varies depending on the type of qubits and the nature of the coupling (photon mediated, direct capacitive or conductive, etc.) between the qubits as well as on the field-qubit coupling strength regime but mostly is modeled as a spin-$1/2$ $XYZ$ Heisenberg exchange interaction \cite{Wendin:2017}. The science of quantum information processing is not only concerned with the fine preparation of such systems in a well defined state but also the controlled time evolution of them while preserving the entanglement among the different parts of the composed system within its coherence time \cite{Nielsen-Chuang2010}. In QIP, the classical data are mapped on the Hilbert space of the processing system, then the time evolution of the system is followed and at the end a readout measurements of the system registers are performed and the classical output is analyzed.

Quantum entanglement is considered to be the physical resource crucially needed for manipulating linear superposition of the quantum states of the different constituents of composite systems to implement the proposed schemes in QIP. However, the inevitable interaction between the quantum system and its environment leads to loss of entanglement in what is known as the decohering process. While gradual loss of entanglement that obeys the half life-time law and evolves to a state of even quite small entanglement can be treated and even reversed using different approaches such as quantum error correction \cite{Shor:1995, Steane:1996}, decoherence free subspace \cite{ Lidar:1998, Kwiat:2000} and quantum measurement reversal \cite{Kim:2012}, 
entanglement sudden death represents a major threat to QIP as the loss takes place very abruptly leading to a state of zero entanglement \cite{Yu:2004,Yu:2008}. The entanglement dynamics in composite systems of interacting quantum systems (qubits) in presence of external magnetic fields coupled to different types of environments in absence of radiation fields have been studied intensively before, where different approaches for creating, enhancing, controlling and protecting entanglement against decoherence and dissipation were investigated and discussed \cite{Khlebnikov:2002, Wang:2006, Huang:2006, Abliz:2006, Tsomokos:2007, Buric:2008, Dubi:2009, Sadiek:2010, Xu:2011, Sahrapour:2013, Sadiek:2013, Alkurtass:2013, Duan:2013, Wu:2014, Sadiek:2016}.

The great progress in developing new systems that are promising candidates for QIP and controllable in cavity or circuit QED sparked huge interest in studying entanglement dynamics and sharing in composite systems containing two-level atoms (qubits) coupled to radiation fields.
The entanglement sharing among the different constituents of a composite system containing two uncoupled atoms interacting with a radiation field was investigated and constrains on the shares were provided \cite{Tessier:2003}. A big turn in the field took place when Yu and Eberly \cite{Yu:2004} showed that the entanglement between two non-interacting, initially entangled, two-level atoms vanishes within a finite period of time, where each atom was coupled to a different environment (cavity), it was called by them, for the first time, entanglement sudden death (ESD).
Latter, the effect of coupling between the two atoms on the revival of the vanished entanglement was studied starting from a particular initial state, where the radiation field was treated as an environment represented by a vacuum state and the master equation of the composite system was solved \cite{Liu:2006,Ficek:2006}. The entanglement sudden death in two initially entangled, uncoupled, atoms under the effect of a noisy classical environment (stochastic magnetic field) was studied both collectively and separately starting from a mixed state \cite{Yu:2006}. More works were devoted to studying the ESD in systems of two uncoupled atoms, each one was in a separate independent cavity, where the effect of different initial types of Bell states was investigated \cite{Yonac:2006, Sainz:2006, Yonac:2007, Sainz:2007}. A double JC model out of resonance with the fields was studied too, where two non-identical, uncoupled atoms were considered in two remote cavities and each atom was coupled to a single mode radiation field in its cavity \cite{Chan:2009}. It was shown that asymmetry can be an advantage for entanglement creation and evolution and the off-resonance condition may, for certain initial states, enhance entanglement transfer (between the atoms and the fields) and prevent ESD.

Recently, the effect of coupling between atoms (qubits) on the system entanglement was brought to the focus of interest, as a result of the newly engineered systems that enjoy strong interactions with each other as we explained earlier, and also due to both its central role in the system dynamics and its practical impact on QIP protocols. The Entanglement dynamics and population difference in a system of two interacting spins have been studied under the effect of coupling to Ohmic and subohmic bosonic environment \cite{Deng:2016}. It was shown that there is a bath-induced spin-spin coupling and the spin-spin entanglement dynamics can be controlled by detuning the coupling to the bosonic bath but depends critically on the initial state of the system. ESD was studied for a system of two identical interacting atoms in a double mode radiation field with frequencies $\omega_1$ and $\omega_2$ at resonance condition, where the atom energy gap $\omega_0$ was such that $\omega_0=\omega_1 + \omega_2$ \cite{Zhang:2007}. An analytical solution for the problem was introduced and it was shown that the time evolution of the entanglement and the ESD depends significantly not only on the initial amount of entanglement in the system but also on the type of initial state. It was demonstrated that ESD was reduced after introducing the dipole-dipole interaction between the two atoms. Also the dipole-dipole interaction was found to enhance the entanglement between the two atoms starting from a W-like initial state as long as the interaction is stronger than the atom field interaction at resonance \cite{Li:2009}. The analytical solution for the same system was presented at off-resonance but for non-interacting atoms and it was shown that the non-zero detuning may enhance the entanglement and suppress the sudden death. An exact analytical solution for the density matrix of two identical interacting atoms coupled to a single mode radiation field at resonance was presented in \cite{Torres:2010}, where the effect of the interplay between the atom-atom interaction and the coupling, at resonance, to the radiation field on both of the entanglement and purity of the system was investigated thoroughly. Recently, a non-linear model for two interacting atoms coupled to a radiation field was introduced by Sanches {\it et al}. \cite{Sanchez:2016}. They considered two types of interaction between the two atoms, dipole-dipole and Ising and represented the radiation field as a coherent superposition of number states. The non-linearity was introduced by introducing and multiplying photon-number-dependent function everywhere times the photon operators in the system Hamiltonian. They diagonalized the interaction Hamiltonian using a basis containing three states and solved Schrodinger equation to obtain the time evolved states at any time $t$ in terms of the basis three-states. Although they started with a Hamiltonian for a generic system where they assumed non-identical atoms out of resonance with the field, but eventually when they came to study specific cases, they assumed identical atoms at resonance condition (zero detuning) with the field, where they provided an exact analytical solution for that case. They studied the time evolution of population inversion, purity of the atomic state and entropy of the radiation field, where they always assumed an initial state that is one of the three states in the implemented basis to be capable of performing the calculations. They carried out numerical calculations to study the entanglement between the two atoms using the concurrence function. Another work studied a non-linear model of two atoms interacting with a radiation field far from resonance where the interaction was considered intensity dependent, but the two atoms were uncoupled \cite{Tavassoly:2018}. Very recently, schemes to avoid ESD in an evolving system of two coupled qubits exposed to a common vacuum environment using Local unitary operations was introduced, where the Lehemberg-Agarwal master equation was implemented under the Markovian approximation \cite{ Chathavalappil:2019}. In a very relevant work, Gywat {\it et al} studied a system of two coupled two-level atoms (qubits) interacting with an off-resonance single mode radiation field \cite{Gywat:2006}. In order to study the system dynamics, they applied a generalized Schrieffer-Wolf transformation to the system Hamiltonian and provided a perturbative analytical solution in the limit of weak interqubit coupling and an exact solution by numerical diagonalization of the Hamiltonian. They demonstrated that the state of the two qubits can be read out using the cavity mode dispersion within the perturbative regime and studied the effect of the interqubit coupling on a cavity-mediated two-qubit gate. In another relevant work, the dynamics of entanglement in a system of two uncoupled spins (qubits) interacting with a single mode radiation field in an optical cavity was studied \cite{Bai:2017}. The time-dependent quantum correlation of Clauser-Horne-Shimony-Holt type was derived and used versus an entanglement measure, concurrence, to test the Bell inequality (BI) violation. It was shown that the interaction with the field induces decoherence and coherence revival that is characterized by the BI violation, where the important role of the field intensity in the decohering process and the inequality test was pointed out.
It is essential to emphasize here that although the pioneering work of Yu and Eberly \cite{Yu:2004} introduced the ESD in the case of an open system, where the ESD took place as result of a dissipative effect (vacuum noise) and also in another work they showed a similar behavior caused by classical noise \cite{Yu:2006}, the same two authors have reported the same phenomenon, ESD, taking place in a closed system. The system contained two uncoupled atoms $(A,B)$, where each one of them is in a separate cavity with a single mode radiation field in each and the two cavities $(a,b)$ are not coupled to each other or to any other environment \cite{Yonac:2006,Yonac:2007}. As they explained in their work, the ESD takes place (zero value of the concurrence $C_{A,B}$ for finite time) in the system, although there is no interactive decoherence, by transferring the entanglement to one or more of the constituting pairs of the composite system: $(a,b)$, $(A,a)$, $(B,b)$, $(A,b)$, $(B,a)$. As a result, the lost entanglement $C_{A,B}$ is gained back within a finite time, which is the reason a collapse-revival behavior of the entanglement is observed. In fact, other authors have reported the same behavior in similar closed systems containing two qubits in either two remote cavities or a single cavity \cite{Sainz:2007,Chan:2009,Zhang:2007,Li:2009}.

In this paper, we consider two identical coupled two-level atoms (qubits) symmetrically interacting with a single-mode quantized radiation field. We present an exact analytical solution for the the time evolution of the system at either resonance or off-resonance (non-zero detuning) interaction between the atoms and the field starting from any initial state of the composite system. The coupling between the two atoms is modeled as a spin 1/2 isotropic XY exchange interaction. As we have discussed before, this system has been treated in several works, in absence of either the coupling between the atoms or the non-zero detuning and was studied in presence of both only in a perturbative way. Our general exact analytical solution provide a mean for studying the different dynamical properties of the system while spanning the whole system parameters space taking into account the interplay among all of them without excluding any. We utilize this solution to study the entanglement dynamics of the system in general and specially  in the case of ESD. Our goal is to test the effect of the coexistence and tuning of atom-atom coupling and off-resonance atom-field interaction, which was not possible before, on the system dynamics and particularly the manipulation of ESD. On the other hand studying this model is not only important for its own sake as a system of coupled localized spins interacting (off-resonance) with a bosonic bath but also as an enlightening step in exploring cavity (circuit) QED, with its crucial impact on QIP schemes as explained earlier. We demonstrate how these two interactions when applied separately or combined can be used to reduce, eliminate or create entanglement sudden death in the system, depending crucially on the initial state. While one or both of the interactions may not be effective individually in treating some cases of ESD, combining them proves to be significantly different. We consider different initial states of the system of special practical interest, which contain maximum, partial or zero entanglement between the two atoms. Also, we show that there is a strong synchronization between the population inversion collapse-revival pattern and the entanglement dynamics at all system parameters combinations. However, only for initial states that may evolve to ESD, the exchange of energy between the field and the atoms enhances the entanglement between the two atoms inducing revival peaks with rapid oscillation, while the population collapse periods synchronize with that of the ESD. Also, the impact of the variation of the radiation field intensity on the ESD duration periods is investigated.

This paper is organized as follows. In Sec. 2, we discuss our model. In Sec. 3, we present our exact analytical solution for the time evolution of the system. We implement our solution to study the dynamics of entanglement and atomic population inversion, starting from different initial sates, in Sec. 4. We conclude in Sec. 5. 
\section{The Model}
\label{The Model}
\begin{figure}[ht]
\centering
\subfigure{\includegraphics[width=12cm]{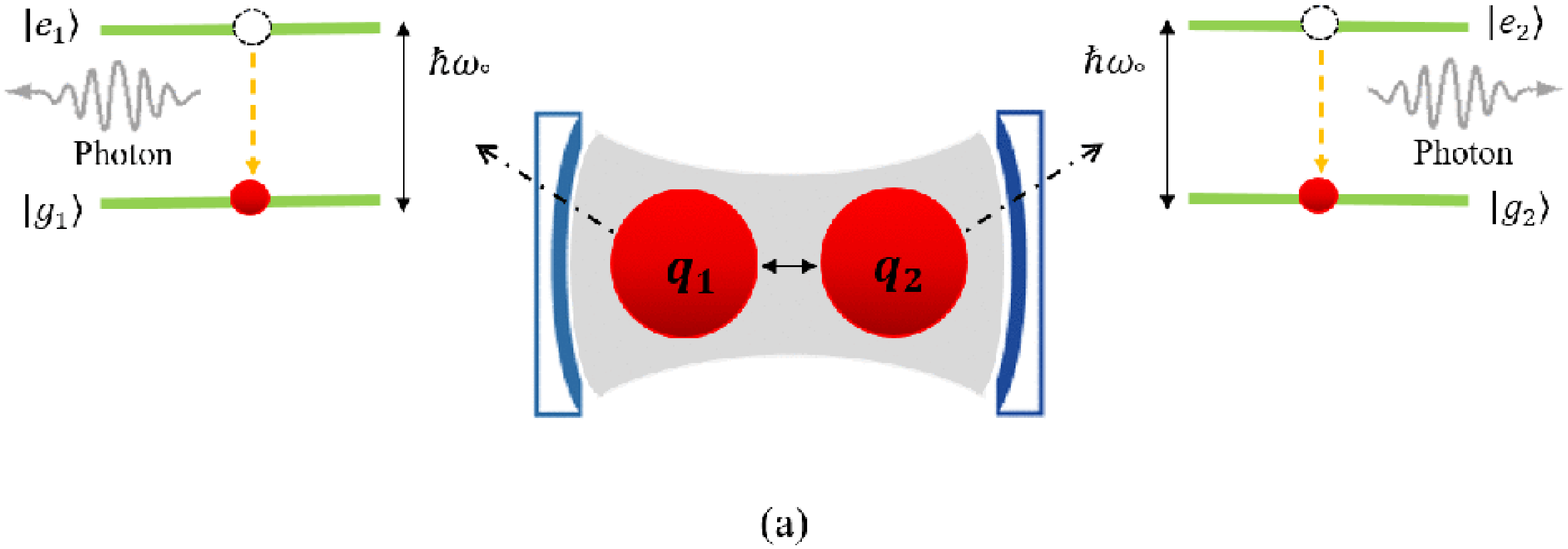}}\\
 \subfigure{\includegraphics[width=12cm]{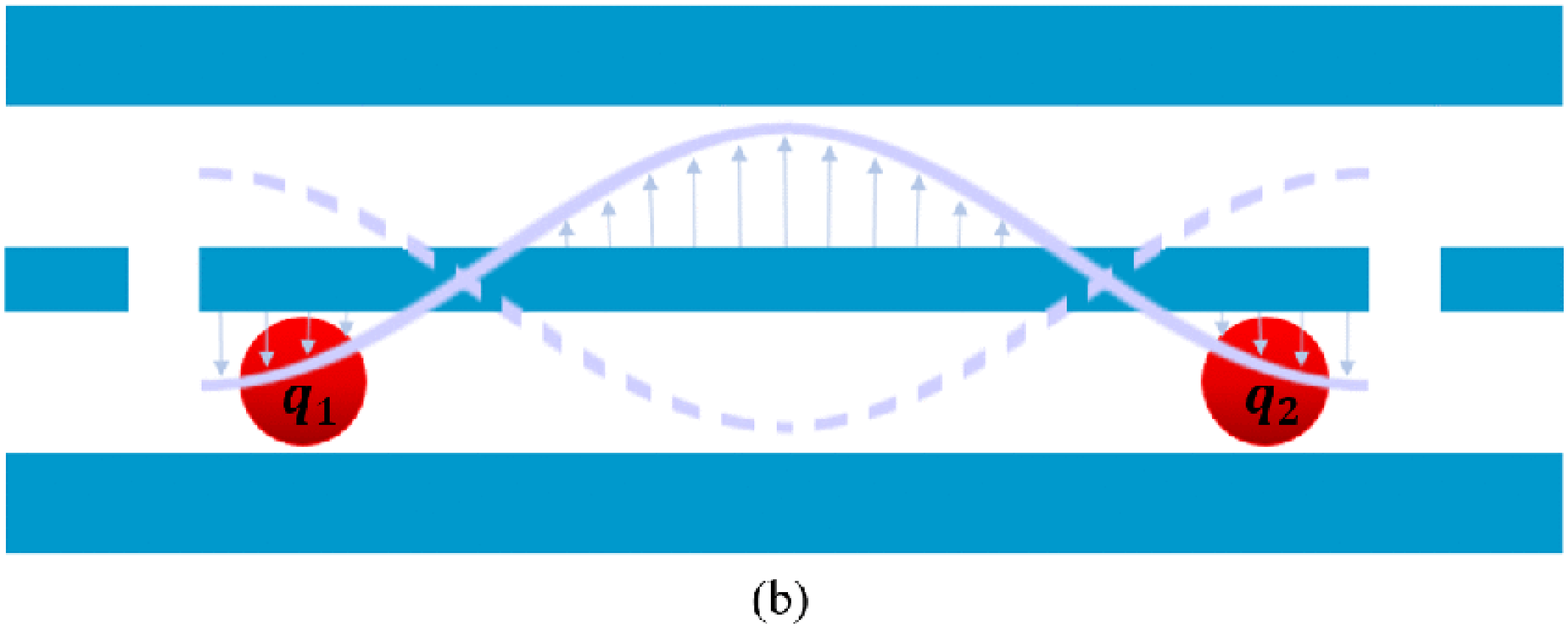}}
\caption{{\protect\footnotesize Schematic diagram of two two-level coupled atoms (qubits) $q_1$ and $q_2$ in (a) an optical cavity or (b) a superconducting microwave resonator.}}
\label{fig1}
\end{figure}

We consider a model of two identical atoms (qubits), each one of them is characterized by two levels: ground $\left|g_{i}\right\rangle$ and excited $\left|e_{i}\right\rangle$, where $i=1,2$ corresponding to the first and second atoms respectively. The two atoms are coupled to the same single-mode quantized radiation field with the same coupling constant $\lambda_1$. The coupling between the two atoms is modeled as an isotropic $XY$ exchange interaction between two spin-1/2 particles with coupling strength $\lambda_2$. This system can be realized in either cavity or circuit QED structures as illustrated in Fig.~\ref{fig1}. The system Hamiltonian assumes the same form as Eq.~(\ref{H_multi-qubit}) except that it is for only two qubits and therefore reduces to
\begin{equation}
\hat{H} = \Omega \; \hat{a}^\dagger \hat{a} \; + \frac{\omega_{\circ}}{2} \sum_{i=1,2} \; \hat{\sigma}^{(i)}_{z}\; + \lambda_1 \sum_{i=1,2} \; (\hat{a}\hat{\sigma}^{(i)}_{+}+\hat{a}^\dagger \hat{\sigma}^{(i)}_{-}) \; + \lambda_2 \; (\hat{\sigma}^{(1)}_{-}\hat{\sigma}^{(2)}_{+}+\hat{\sigma}^{(1)}_{+}\hat{\sigma}^{(2)}_{-})\;.
\label{eq:H}
\end{equation}
As in Eq.~(\ref{H_multi-qubit}), the first and second terms in the Hamiltonian represent the free quantized radiation field and the non-interacting two atoms while the third and fourth terms represent the atom-field and atom-atom interactions respectively. $\Omega$ and $\omega_{\circ}$ are the frequencies of the single-mode radiation field and the quantum system transition respectively, $\hat{a}^\dagger$, $\hat{a}$, $\hat{\sigma}^{(i)}_{\pm}$ and $\hat{\sigma}^{i}_{z}$ have the same meaning and roles as in Eq.~(\ref{H_multi-qubit}).
\\
Using the Heisenberg equation of motion, which for any operator $\hat{O}$, where $\hbar=1$, reads 
\begin{equation}
\frac{d\hat{O}}{dt}=\textrm{-i}[\hat{Q},\hat{H}] + \frac{\partial \hat{O}}{\partial t},
\label{eq:Heisenberg_eqn_of motion}
\end{equation}
one obtains the following equations for the field and atom operators
\begin{eqnarray}
\nonumber \frac{d\hat{a}}{dt} &=& -\textrm{i} \Omega \hat{a} - \textrm{i} \lambda_{1} (\hat{\sigma}^{(1)}_{-}+\hat{\sigma}^{(2)}_{-}),\\
\nonumber \frac{d\hat{\sigma}^{(1)}_{-}}{dt} &=& -\textrm{i} \omega_{\circ} \hat{\sigma}^{(1)}_{-} + \textrm{i} \lambda_{1} \hat{a} \hat{\sigma}^{(1)}_{z} +  \textrm{i} \lambda_{2} \hat{\sigma}^{(1)}_{z} \hat{\sigma}^{(2)}_{-},\\
\nonumber \frac{d\hat{\sigma}^{(2)}_{-}}{dt} &=& -\textrm{i} \omega_{\circ} \hat{\sigma}^{(2)}_{-} + \textrm{i} \lambda_{1} \hat{a} \hat{\sigma}^{(2)}_{z} +  \textrm{i} \lambda_{2} \hat{\sigma}^{(1)}_{-}\hat{\sigma}^{(2)}_{z},\\
\nonumber \frac{d\hat{\sigma}^{(1)}_{z}}{dt} &=& 2\textrm{i} \lambda_{1} (\hat{a}^\dagger \hat{\sigma}^{(1)}_{-}-\hat{a}\hat{\sigma}^{(1)}_{+})+ 2\textrm{i} \lambda_{2} (\hat{\sigma}^{(1)}_{-}\hat{\sigma}^{(2)}_{+}-\hat{\sigma}^{(1)}_{+}\hat{\sigma}^{(2)}_{-}),\\
\frac{d\hat{\sigma}^{(2)}_{z}}{dt} &=& 2\textrm{i} \lambda_{1} (\hat{a}^\dagger \hat{\sigma}^{(2)}_{-}-\hat{a}\hat{\sigma}^{(2)}_{+})+ 2\textrm{i} \lambda_{2} (\hat{\sigma}^{(1)}_{+}\hat{\sigma}^{(2)}_{-}-\hat{\sigma}^{(1)}_{-}\hat{\sigma}^{(2)}_{+}),
\label{eq:opertors_dynamics}
\end{eqnarray}
Assuming that initially the atoms are in a pure state and the field is in the coherent state, the wave function of the composite system at $t=0$ can be written as
\begin{equation}
\vert\psi(0)\rangle = [a\; \vert e_{1},e_{2}\rangle + b \; \vert e_{1},g_{2}\rangle + c \; \vert g_{1},e_{2}\rangle + d \; \vert g_{1},g_{2}\rangle]\otimes \vert\alpha\rangle,
\label{eq:8}
\end{equation}
where $a,b,c$ and $d$, are arbitrary complex quantities that satisfy the condition 
\begin{equation}
\vert a \vert^2 + \vert b \vert^2 +\vert c \vert^2 +  \vert d \vert^2 =1,
\label{eq:9}
\end{equation}
and $\vert\alpha\rangle$ is the coherent state defined as
\begin{equation}
\vert\alpha\rangle=\sum_{n} Q_{n} \vert n \rangle;   \qquad Q_{n}=\frac{\alpha^{n}}{\sqrt{ n !}} \exp\left(-\frac{\vert \alpha \vert^2}{2}\right),
\label{eq:10}
\end{equation}
where $\vert \alpha \vert^2=\overline{n}$ is the mean photon number and $\vert n \rangle$ are the photon number states, which satisfy the relations: $\hat{a}^{\dagger} \vert n \rangle =\sqrt{n+1}\vert n+1 \rangle$ and $\hat{a} \vert n+1 \rangle = \sqrt{n+1} \vert n \rangle$. The wave function at any time $t$ latter can be written as
\begin{eqnarray}
\nonumber \vert\psi(t)\rangle &=& \sum_{n} [ A_{n}(t)\vert e_{1},e_{2},n\rangle + B_{n+1}(t)\vert e_{1},g_{2},n+1\rangle + C_{n+1}(t) \vert g_{1},e_{2},n+1\rangle \\
&& + D_{n+2}(t)\vert g_{1},g_{2},n+2\rangle ],
\label{eq:psi_t}
\end{eqnarray}
where $ \vert e_{1},e_{2},n\rangle $ is the state in which the two atoms are in excited state and the field has $n$ photons, $ \vert e_{1},g_{2},n+1\rangle $ is the state in which the first one is in the excited state and the second is in the ground state and the field has $n+1$ photons and so on. The states of the quantum system satisfy the relations $\hat{\sigma}_{+}^{(i)} \vert g_i \rangle = \vert e_i \rangle$, $\hat{\sigma}_{-}^{(i)} \vert g_i \rangle =0$, $\hat{\sigma}_{+}^{(i)} \vert e_i \rangle = 0$ and $\hat{\sigma}_{-}^{(i)} \vert e_i \rangle =  \vert g_i \rangle$. The time-dependent coefficients $A_{n}(t), B_{n+1}(t), C_{n+1}(t)$ and $D_{n+2}(t)$ can be obtained by solving the Schr{\"o}dinger equation of the composite system, which will be discussed in the next section. Once we obtain the system wave function $\vert\psi(t)\rangle$, we can calculate the composite system density matrix $\hat{\rho}(t)=\vert\psi(t)\rangle\langle\psi(t)\vert$. The reduced density matrix of the two atoms, $\hat{\rho}_{\textrm{red}}(t)$, can be obtained by tracing out the field 
\begin{equation}
\hat{\rho}_{\textrm{red}}(t)=\textrm{Tr}_{\textrm{field}}\; \hat{\rho}(t)= \sum_{l} \langle l \vert \psi(t)\rangle \langle \psi(t) \vert l \rangle.
\label{eq:qs_rdm}
\end{equation}
\section{The Analytical Solution}
\label{The Analytical Solution}
We devote this section to solve the Schr{\"o}dinger equation of the system and provide an exact analytical expression for the time-dependent coefficients of the system wave function. We start by rewriting the Hamiltonian (Eq.~(\ref{eq:H})) as 
\begin{equation}
\hat{H}=\hat{H}_{\circ}+\hat{H}_{int}\;,
\label{eq:14}
\end{equation}
where 
\begin{equation}
\hat{H}_{\circ} = \Omega \; \hat{N} + \frac{\Delta}{2} \sum_{i=1,2} \; \hat{\sigma}^{(i)}_{z}\;,
\label{eq:15}
\end{equation} 
\begin{equation}
\hat{H}_{int}= \lambda_1 \sum_{i=1,2} \; (\hat{a}\hat{\sigma}^{(i)}_{+}+\hat{a}^\dagger \hat{\sigma}^{(i)}_{-})\: + \lambda_2 \; (\hat{\sigma}^{(1)}_{-}\hat{\sigma}^{(2)}_{+}+\hat{\sigma}^{(1)}_{+}\hat{\sigma}^{(2)}_{-}),
\label{eq:16}
\end{equation}
and
\begin{equation}
\hat{N}= \hat{a}^\dagger \hat{a} + \frac{1}{2} \sum_{i=1,2} \; \hat{\sigma}^{(i)}_{z}\;, 
\label{eq:5}
\end{equation}
where $\Delta=\omega_{\circ}-\Omega$ is the detuning parameter. 
Using Eqs. \ref{eq:Heisenberg_eqn_of motion} and \ref{eq:opertors_dynamics}, one can show that $\hat{N}$, which represents the total number of excitations in the system, is a constant of motion, which justifies the use of its eigenstates as a basis for the expansion of the system wavefunction in Eq. (\ref{eq:psi_t}).
It is more convenient to work in the interaction picture where we define $\hat{V}_I=\hat{U} \hat{H}_{int} \hat{U}^{\dagger}$ with $\hat{U}=e^{\textrm{i}\hat{H}_{\circ}t}$. As a result, we obtain
\begin{equation}
\hat{V}_{I}(t)= \; \lambda_1 \sum_{i=1,2}\; (\hat{a}\;e^{\textrm{i}\Delta t} \hat{\sigma}^{(i)}_{+} + \hat{a}^\dagger \; e^{-\textrm{i}\Delta t} \hat{\sigma}^{(i)}_{-})+ \hbar \: \lambda_2 ( \hat{\sigma}^{(1)}_{-}\hat{\sigma}^{(2)}_{+}\\
+ \hat{\sigma}^{(1)}_{+}\hat{\sigma}^{(2)}_{-})\;.
\label{eq:V_int}
\end{equation}
Now, substituting $\vert\psi(t)\rangle$ and $V_I(t)$ into Schr{\"o}dinger equation
\begin{equation}
\textrm{i} \; \frac{\partial}{\partial t}\vert\psi(t)\rangle = \hat{V}_I(t)\vert\psi(t)\rangle, 
\label{eq:Sch_eqn}
\end{equation}
it yields a system of coupled differential equations
\begin{eqnarray}
\nonumber \textrm{i}\dot{A}_{n}(t) &=& \alpha\; e^{\textrm{i}\Delta t} \;(B_{n+1}(t) + C_{n+1}(t)),\\
\nonumber \textrm{i}\dot{B}_{n+1}(t) &=& \alpha \;e^{-\textrm{i} \Delta t }\;A_{n}(t) + \beta \;e^{\textrm{i}\Delta t}\;D_{n+2}(t) + \lambda_{2}\;C_{n+1}(t),\\
\nonumber \textrm{i}\dot{C}_{n+1}(t) &=& \alpha\;e^{-\textrm{i} \Delta t }\;A_{n}(t) + \beta \;e^{\textrm{i}\Delta t}\;D_{n+2}(t) + \lambda_{2}\;B_{n+1}(t),\\
\textrm{i}\dot{D}_{n+2}(t)&=& \beta\; e^{-\textrm{i}\Delta t}\;(B_{n+1}(t) + C_{n+1}(t)),
\label{eq:DE_sys1}
\end{eqnarray}
where $\alpha=\lambda_{1}\sqrt{n+1}$ and $\beta=\lambda_{1}\sqrt{n+2}$. Substituting $ K(t)= B_{n+1}(t)+C_{n+1}(t)$, Eqs.~(\ref{eq:DE_sys1}) simplify to
\begin{eqnarray}
\nonumber \textrm{i}\dot{A}_{n}(t) &=& \alpha K(t)\;e^{\textrm{i}\Delta t},\\
\nonumber \textrm{i}\dot{D}_{n+2}(t)&=& \beta K(t)\;e^{-\textrm{i}\Delta t},\\
\textrm{i}\dot K(t) &=& 2 \alpha\;e^{-\textrm{i} \Delta t}\;A_{n}(t) + 2\beta e^{\textrm{i}\Delta t} D_{n+2}(t) + \lambda_{2} K(t),
\label{eq:DE_sys2}
\end{eqnarray}
which after some calculations becomes
\begin{equation}
\dddot K(t) + \textrm{i}\lambda_{2}\ddot K(t) + [2(\alpha^{2}+\beta^{2})+\Delta^{2}]\;\dot K(t) - \textrm{i}[2\Delta(\alpha^{2}-\beta^{2})-\lambda_{2}\Delta^{2}]\;K(t)=0,
\label{eq:3rd_order_DE}
\end{equation}
with a solution
\begin{equation}
K(t)= \sum^{3}_{j=1}\delta_{j} e^{m_{j}t},
\label{eq:23}
\end{equation}
where
\begin{eqnarray}
\nonumber \delta_{1}&=&(B_{n+1}(0)+C_{n+1}(0))-(\delta_{2}+\delta_{3}),\\
\nonumber \delta_{2}&=&\frac{1}{(m_{1}-m_{2})(m_{3}-m_{2})} \lbrace 2\alpha A_{n}(0)[\textrm{i}(m_{1}+m_{3})-\lambda_{2}-\Delta]+2\beta D_{n+2}(0) [\textrm{i}(m_{1}+m_{3})\\
\nonumber &&- \lambda_{2}+\Delta]+ [\textrm{i}(m_{1}+m_{3})(\lambda_{2}-\textrm{i}m_{1})-2(\alpha^{2}+\beta^{2})-\lambda^{2}_{2}-m_{1}^{2}]\\
\nonumber && \times(B_{n+1}(0)+C_{n+1}(0))\rbrace, \\
\nonumber \delta_{3}&=&\frac{1}{(m_{1}-m_{3})(m_{2}- m_{3})}\lbrace 2\alpha A_{n}(0)[\textrm{i}(m_{1}+m_{2})-\lambda_{2}-\Delta]+2\beta D_{n+2}(0)[\textrm{i}(m_{1}+m_{2})\\
\nonumber &&-\lambda_{2}+\Delta]+[\textrm{i}(m_{1}+m_{2})(\lambda_{2}-\textrm{i}m_{1})-2(\alpha^{2}+\beta^{2})-\lambda_{2}^{2}-m_{1}^{2}]\\
&& \times(B_{n+1}(0)+C_{n+1}(0))\rbrace,
\label{eq:24}
\end{eqnarray}
and 
\begin{eqnarray}
\nonumber m_{1}&=&(v_{1}+v_{2})-\textrm{i}\frac{\lambda_{2}}{3},\\
\nonumber m_{2}&=&-\frac{v_{1}+v_{2}}{2}+\textrm{i} \frac{\sqrt{3}}{2}(v_{1}-v_{2})-\textrm{i}\frac{\lambda_{2}}{3},\\
m_{3}&=&-\frac{v_{1}+v_{2}}{2}-\textrm{i} \frac{\sqrt{3}}{2}(v_{1}-v_{2})-\textrm{i}\frac{\lambda_{2}}{3},
\label{eq:ms}
\end{eqnarray}
where
\begin{equation}
v_{1}=[ -\frac{\mu}{2}+(\frac{\mu^{2}}{4}+\frac{\eta^{3}}{27})^\frac{1}{2}]^\frac{1}{3}; \qquad
v_{2}=[-\frac{\mu}{2}-(\frac{\mu^{2}}{4}+\frac{\eta^{3}}{27})^\frac{1}{2}]^\frac{1}{3},
\label{eq:v1_v2}
\end{equation}
and 
\begin{equation}
\mu=-\frac{\textrm{i}}{27} [2\lambda_{2}^{3}+18\lambda_{2}(\alpha^{2}+\beta^{2}-\Delta^2)+54 \Delta (\alpha^{2}-\beta^{2})],
\label{eq:mu}
\end{equation}
\begin{equation}
\eta=\frac{1}{3}[6(\alpha^{2}+\beta^{2})+3\Delta^2+\lambda_{2}^{2}]\;.
\label{eq:eta}
\end{equation}
Finally, the solution of the set of differential equations~(\ref{eq:DE_sys1}) takes the form
\begin{eqnarray}
\nonumber A_{n}(t) &=& A_{n}(0)-\textrm{i}\alpha \sum^{3}_{j=1}[\frac{\delta_{j}}{m_{j}+\textrm{i}\Delta}( e^{(m_{j}+\textrm{i}\Delta )t}-1)],\\
\nonumber B_{n+1}(t)&=&\frac{1}{2}[(B_{n+1}(0)-C_{n+1}(0))e^{\textrm{i}\lambda_{2}t}+\sum^{3}_{j=1}\delta_{j} e^{m_{j}t}],\\
\nonumber C_{n+1}(t)&=&\frac{1}{2}[(C_{n+1}(0)-B_{n+1}(0))e^{\textrm{i}\lambda_{2}t}+\sum^{3}_{j=1}\delta_{j} e^{m_{j}t}],\\
D_{n+2}(t)&=& D_{n+2}(0)-\textrm{i}\beta \sum^{3}_{j=1}[\frac{\delta_{j}}{m_{j}-\textrm{i}\Delta}( e^{(m_{j}-\textrm{i}\Delta )t}-1)],
\label{eq:coef_soln}
\end{eqnarray}
where the initial values of the coefficients are given by
\begin{eqnarray}
A_{n}(0)=Q_{n}\;a, \;\;\; B_{n+1}(0)=Q_{n+1}\;b,\;\;\; C_{n+1}(0)=Q_{n+1}\;c,\;\;\; D_{n+2}(0)=Q_{n+2}\;d.
\label{eq:init_values}
\end{eqnarray}

As can be noticed, for Eqs.~(\ref{eq:coef_soln}) to represent an acceptable physical solution, the parameters $m_1$, $m_2$ and $m_3$ in the exponents can have only either negative or imaginary values, otherwise the coefficients will blow up with time. This restriction causes certain roots of $v_1$ and $v_2$ in Eqs.~(\ref{eq:v1_v2}) to be appropriate for the solution whereas the others represent a non-physical solution. In fact, each one of the two quantities $v_1$ and $v_2$ will have three, generally complex, roots. Therefore $v_{1}$ and $ v_{2}$ defined by Eqs.~(\ref{eq:v1_v2}) have nine possible combinations, only six of them lead to physically acceptable solution. Nevertheless, very fortunately these six combinations enable us to span the whole parameter space of the system. Finally the reduced density matrix of the two atoms defined by Eq.~(\ref{eq:qs_rdm}) can be obtained, utilizing that $\rho^{\dagger}=\rho$, as
\begin{equation}
\rho_{red}= \displaystyle \sum_{n=0}^{\infty} \left( \begin{matrix}
\vert A_{n} \vert^2 & A_{n+1}B_{n+1}^{*} & A_{n+1}C_{n+1}^{*} & A_{n+2}D_{n+2}^{*}\\
B_{n+1}A_{n+1}^{*} & \vert B_{n+1} \vert^2 & B_{n+1}C_{n+1}^{*} & B_{n+2}D_{n+2}^{*}\\
C_{n+1}A_{n+1}^{*} & C_{n+1}B_{n+1}^{*} & \vert C_{n+1} \vert^2 & C_{n+2}D_{n+2}^{*}\\
D_{n+2}A_{n+2}^{*} & D_{n+2}B_{n+2}^{*} & D_{n+2}C_{n+2}^{*} &\vert D_{n+2} \vert^2
\end{matrix} \right),
\label{qs_rdm}
\end{equation}
\section{Dynamics of entanglement and atomic population inversion}
In this section we implement our exact solution to study the dynamics of the bipartite entanglement between the two atoms and the atomic population inversion starting from different initial states of particular interest. For convenience, we set $\hbar=1$, $\lambda_1=1$ and set represent the other parameters ($\lambda_2$ and $\Delta$) in units of $\lambda_1$. The entanglement between the two quantum system can be quantified with the help of the concurrence function $C(\rho_{\textrm{red}})$ as proposed by Wootters~\cite{Wootters:1998}, which is related to the entanglement of formation $E_{f}$ through the formula
\begin{equation}
E_{f}(\rho_{\textrm{red}})= \mathcal{E}(C(\rho_{\textrm{red}})),
\label{eq:48}
\end{equation}
where $\mathcal{E}$ is defined as
\begin{equation}
\mathcal{E}(C(\rho_{\textrm{red}}))= h\left(\frac{1+ \sqrt{1-C^{2}(\rho_{\textrm{red}})}}{2} \right),
\label{eq:49}
\end{equation}
$h$ is the Shannon entropy function
\begin{equation}
h(x)=-x \log_{2} x - (1-x)\log_{2} (1-x),  
\label{eq:50}
\end{equation}
and the concurrence can by calculated from
\begin{equation}
C(\rho_{\textrm{red}})= \max\;[0,\varepsilon_{1}-\varepsilon_{2}-\varepsilon_{3}-\varepsilon_{4}],
\label{eq:51}
\end{equation}
The $\varepsilon_{i}$ arranged in decreasing order are the square root of the four eigenvalues of the non-Hermitian matrix 
\begin{equation}
R\equiv \rho_{\textrm{red}}\tilde{\rho}_{\textrm{red}},
\label{eq:52}
\end{equation}
Where $\tilde{\rho}_{\textrm{red}}$ is the spin flipped state defined as
\begin{equation}
\tilde{\rho}_{\textrm{red}}=(\hat{\sigma}_{y}\otimes\hat{\sigma}_{y})\rho^{*}_{\textrm{red}}(\hat{\sigma}_{y}\otimes\hat{\sigma}_{y}),
\label{eq:53}
\end{equation} 
Here $\rho^{*}_{\textrm{red}}$ is the complex conjugate of $\rho_{\textrm{red}}$ and $\hat{\sigma}_{y}$ is the Pauli spin matrix in the $y$ direction.
Both of  $C(\rho_{\textrm{red}})$ and $E_{f}(\rho_{\textrm{red}})$ go from $0$ for a separable state to $1$ for a maximally entangled state.

Atomic population inversion is defined as the expectation value of the operator $\hat{\sigma}_{z}$ or the difference between the probabilities of finding the atom in its excited state and ground state. To investigate the atomic inversion we first calculate the reduced density matrix of any one of the two identical atoms, say the first, $\hat{\rho}_{1}(t)$ by tracing out the other one in the two atoms reduced density matrix $\hat{\rho}_{red}$ (Eq.~\ref{qs_rdm}), which leads to 
\begin{equation}
\hat{\rho}_{1}(t)=Tr_{q_2}\;\hat{\rho}_{red}(t) =\displaystyle  \left(\begin{array}{cc}
\rho_{11} & \rho_{12} \\
\rho_{21} & \rho_{22}
\end{array}\right),
\label{eq:33}
\end{equation} 
where
\begin{eqnarray}
\nonumber \rho_{11}(t)&=&\sum_{n=0}^{\infty} \vert A_{n}(t) \vert^2 +  \vert B_{n+1}(t) \vert^2, \\
\nonumber \rho_{22}(t)&=& \sum_{n=0}^{\infty}\vert C_{n+1}(t) \vert^2 + \vert D_{n+2}(t) \vert^2, \\
\rho_{12}(t)&=& \rho^{*}_{21}(t)= \sum_{n=0}^{\infty} A_{n+1}(t)C_{n+1}^{*}(t) + B_{n+2}(t)D_{n+2}^{*}(t).
\label{eq:rho_elements}
\end{eqnarray}
Therefore, for the first atom 
\begin{eqnarray}
\nonumber\langle \hat{\sigma}_{z}(t) \rangle &=& Tr [\hat{\rho}_{1}(t)\hat{\sigma}_{z}]\\
&=& \sum_{n=0}^{\infty} \vert A_{n}(t) \vert^2 +  \vert B_{n+1}(t)\vert^2 - \vert C_{n+1}(t) \vert^2 - \vert D_{n+2}(t) \vert^2.
\label{eq:sigma_z}
\end{eqnarray}
\subsection{Maximally entangled initial (Bell) states}
\begin{figure}[htbp]
\centering
\begin{subfigure}\centering\includegraphics[width=6.5cm]{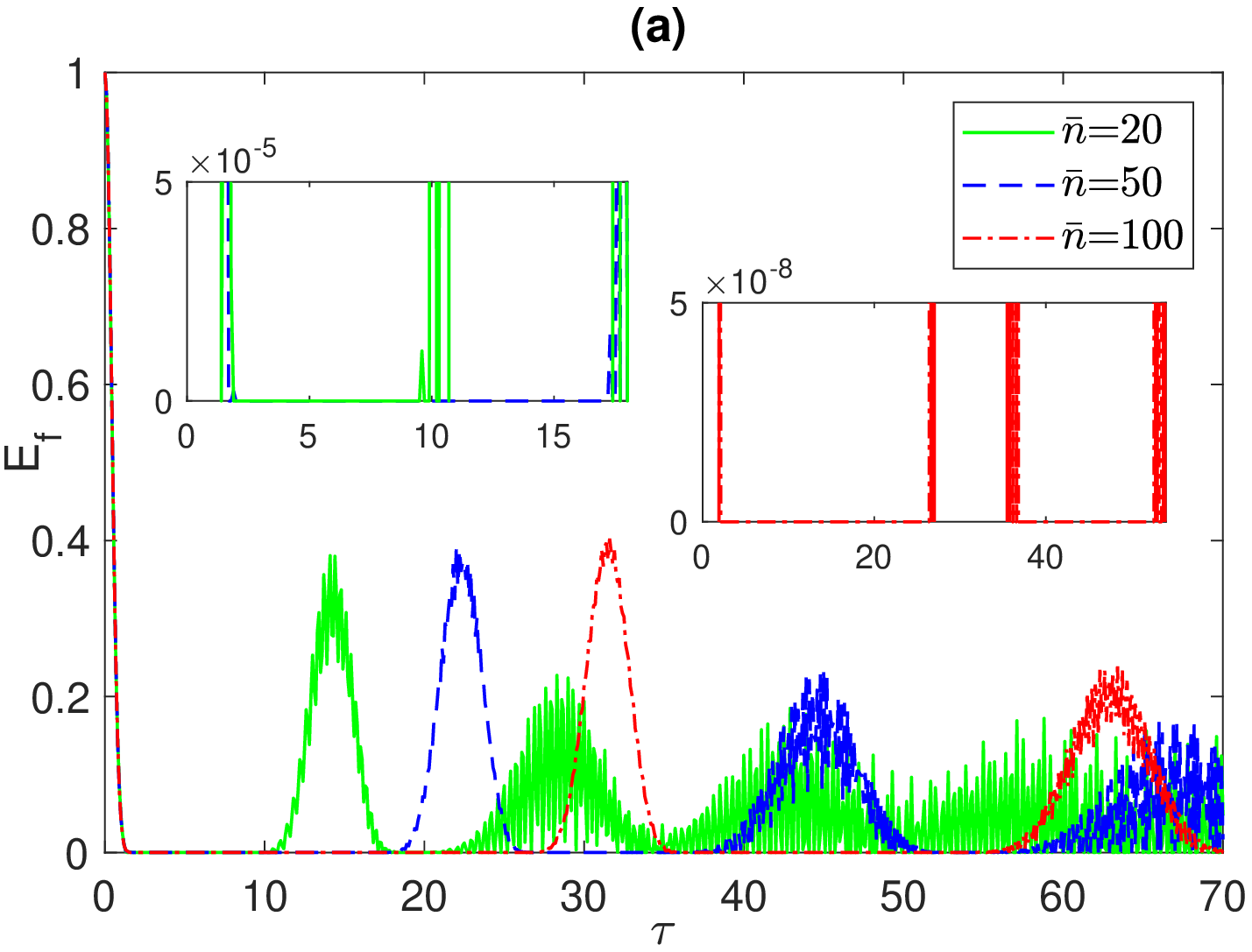}\end{subfigure}
\begin{subfigure}\centering\includegraphics[width=6.7cm]{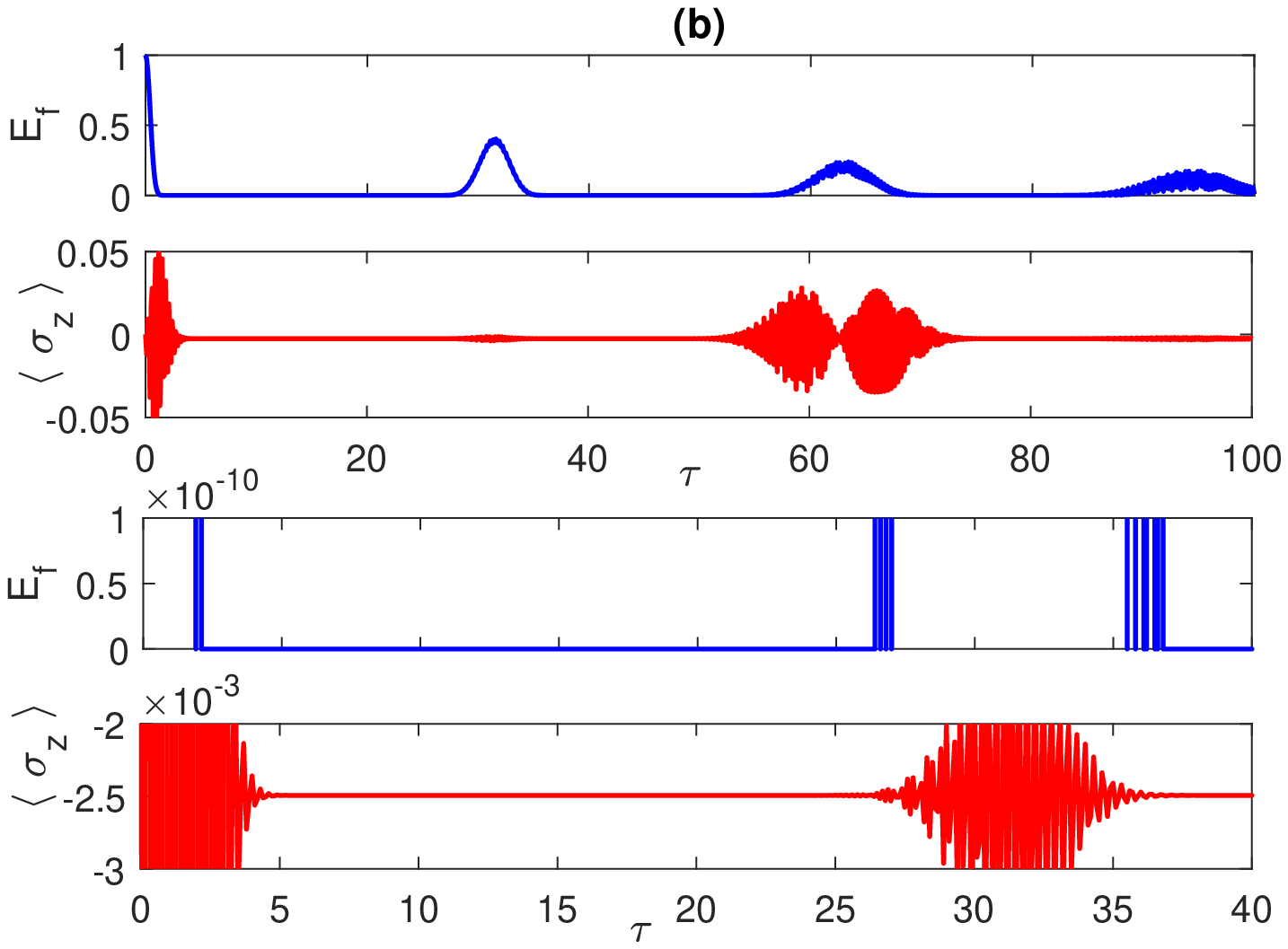}\end{subfigure}
\begin{subfigure}\centering\includegraphics[width=6.6cm]{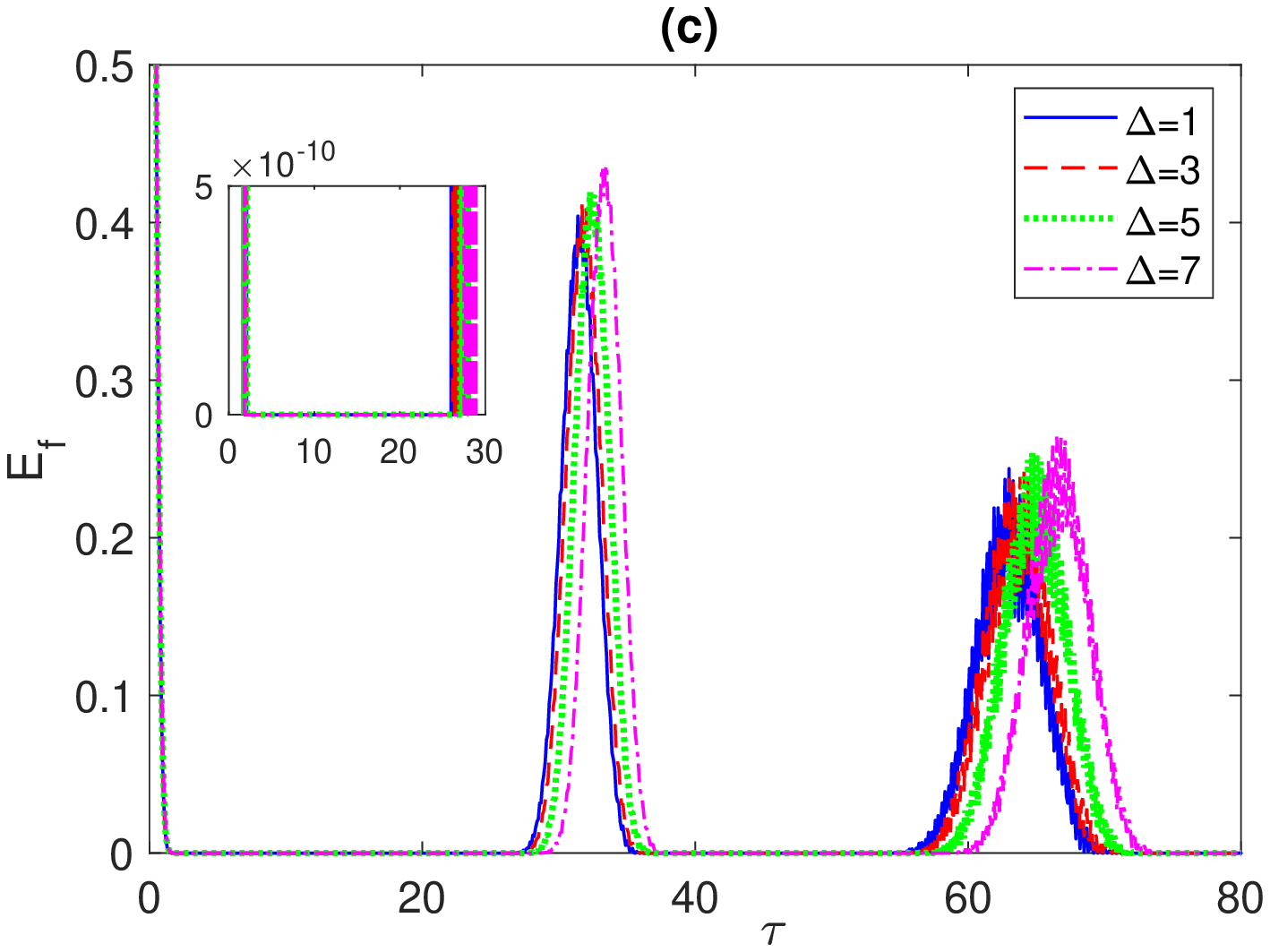}\end{subfigure}
\begin{subfigure}\centering\includegraphics[width=6.6cm]{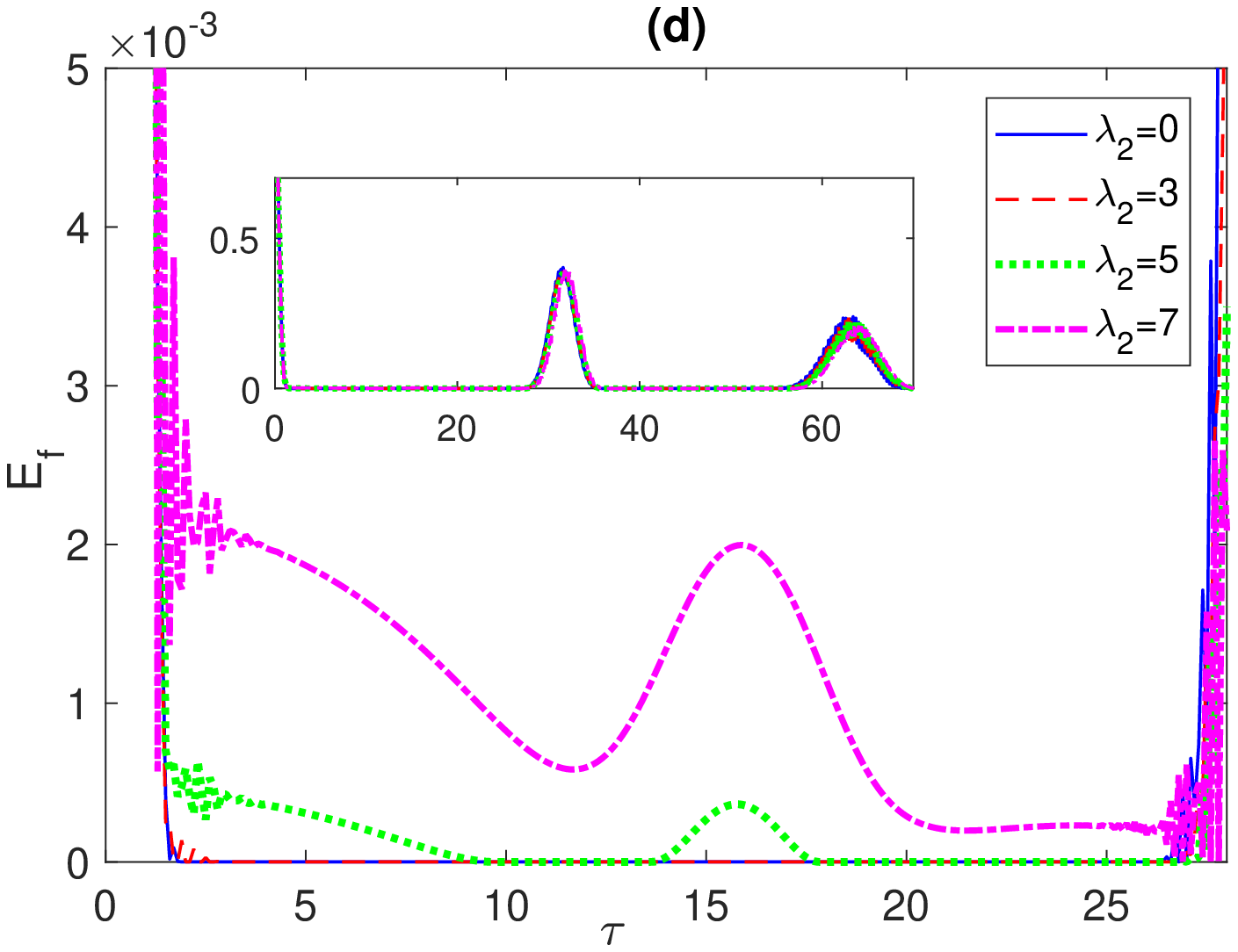}\end{subfigure}
\caption{{
Entanglement $E_f$ and population inversion $\langle \sigma_z \rangle$ versus the scaled time $\tau=\lambda_1 t$ with the two atoms are initially  in a correlated Bell state $\psi_{Bc}=(\vert e_{1}\rangle \vert e_{2}\rangle+\vert g_{1}\rangle \vert g_{2}\rangle)/\sqrt{2}$ and the field is in a coherent state: 
(a) $E_f$ versus $\tau$ for $\lambda_2=0$, $\Delta=0$ and various values of the mean number of photons;
(b) $E_f$ and $\langle \sigma_z \rangle$ versus $\tau$ for $\lambda_2=0$, $\Delta=0$ and $\bar{n}=100$; 
(c) $E_f$ versus $\tau$ for $\lambda_2=0$, $\bar{n}=100$ and various values of $\Delta$, and 
(d) $E_f$ versus $\tau$ for $\Delta=0$, $\bar{n}=100$ and various values of $\lambda_2$.}}
\label{fig2}
\end{figure}

In Fig.~\ref{fig2}, we explore the dynamics of entanglement and population inversion, in terms of the scaled time $\tau=\lambda_1 t$, starting from a correlated initial Bell state $\psi_{Bc}=(\vert e_{1}\rangle \vert e_{2}\rangle+\vert g_{1}\rangle \vert g_{2}\rangle)/\sqrt{2}\;$ with the radiation field is in a coherent state. Starting from such an initial state the system shows ESD, where the entanglement changes abruptly from a non-zero to an exact zero value, which is illustrated in the different panels of the figure. In Fig.~\ref{fig2}(a), we test the effect of the field intensity, by changing the average number of photons $\bar{n}$, on the entanglement dynamics and ESD time intervals for uncoupled atoms at resonance with the radiation field. As can be noticed, by increasing the intensity of the radiation field from $\bar{n}=20$ to $50$ and then $100$, the sudden death interval time increases considerably. 
The inset plots of Fig.~\ref{fig2}(a) show, at a magnified scale, the sharp transition in the entanglement from a finite value to zero, at different field intensity values ($\bar{n} = 20,\; 50$ and $100$). Also, they show the change in the ESD time interval as $\bar{n}$ changes. In the forthcoming discussion, we set $\bar{n}=100$ everywhere except when otherwise is mentioned explicitly. 
The dynamics of entanglement and population inversion is depicted in Fig.~\ref{fig2}(b) for uncoupled atoms at resonance with the field. In the upper two panels, we compare the entanglement dynamics and the atomic population within the interval $0 \leq \tau \leq 100$, which shows three entanglement revival peaks. In the lower two panels, we compare them again but after zooming into a smaller time interval $0 \leq \tau \leq 40$ and much smaller ranges of $E_f$ and $\langle \hat{\sigma}_{z} \rangle$, so we can focus on the first entanglement revival peak and the corresponding atomic population dynamics. As can be noticed in the upper most panel, the two-atoms start in an initial state where they are maximally entangled with each other but abruptly they lose their entanglement, showing ESD, and maintain this state for a finite period of time. However, they gain entanglement back with a revival peak as shown, at $\tau \sim 27$, which as we pointed out in our introduction is due to the fact that the loss of entanglement is not due to a dissipative effect but a transfer of entanglement to the atom-field subsystems, which were initially disentangled. In the second upper panel, the atomic population shows the usual collapse revival pattern, although it doesn't collapse to zero but a constant value. This behavior of the entanglement and atomic population is repeated periodically. Clearly, the collapse periods coincide with the zero entanglement intervals, while the population revivals temporally coincide with the entanglement revivals. This behavior can be better recognized in the lower two panels where one can see  
that for the atoms entanglement to drop to zero a rapid oscillation of the atomic population takes place at the same time, indicating an exchange of energy between the atoms and the field is taking place. Then the zero entanglement state is maintained for a finite time before an entanglement revival occurs, where the entanglement increases from zero to a peak then back to zero, accompanied by a rapid oscillation of the atomic population that starts and finishes within the same time interval. Obviously, the entanglement between the two atoms is gained back from the atom-field subsystems before getting lost to them back through exchange of energy between the atoms and the field. In Fig.~\ref{fig2}(c), the two uncoupled atoms are considered at different detuning parameter values, $\Delta=1, 3, 5$ and 7. The non-zero detuning does not remove or affect the ESD except for increasing the entanglement death intervals slightly as shown. The inset plot in Fig.~\ref{fig2}(c) focuses on the first death period showing the sharp transition from non-zero entanglement to zero and back to non-zero value. Also, it shows the small shift in the entanglement death interval as $\Delta$ increases.  
The effect of the atom-atom coupling at zero detuning is illustrated in Fig.~\ref{fig2}(d), where a small coupling value, $\lambda_2=3$, has no effect on the ESD (dashed red line), while a higher value, 5, partially eliminates the ESD (dotted green line). Further increase of the coupling to $\lambda_2=7$ completely eliminates the ESD (dash dotted violet line). The inset plot in Fig.~\ref{fig2}(d) shows the entanglement collapse-revival pattern over a longer time interval in presence of atomic coupling.   
\begin{figure}[htbp]
\centering
\begin{subfigure}\centering\includegraphics[width=6.6cm]{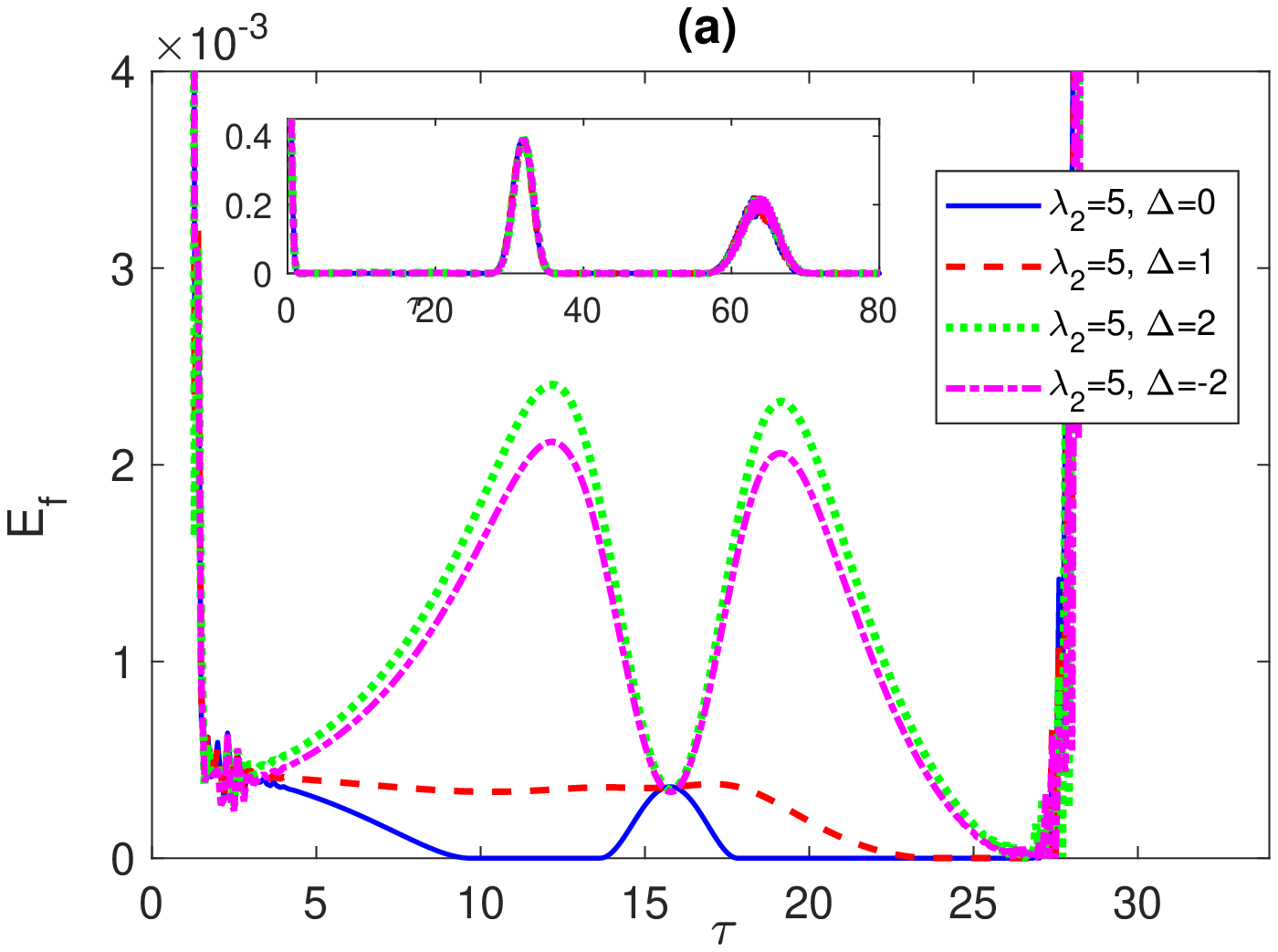}\end{subfigure}
\begin{subfigure}\centering\includegraphics[width=6.6cm]{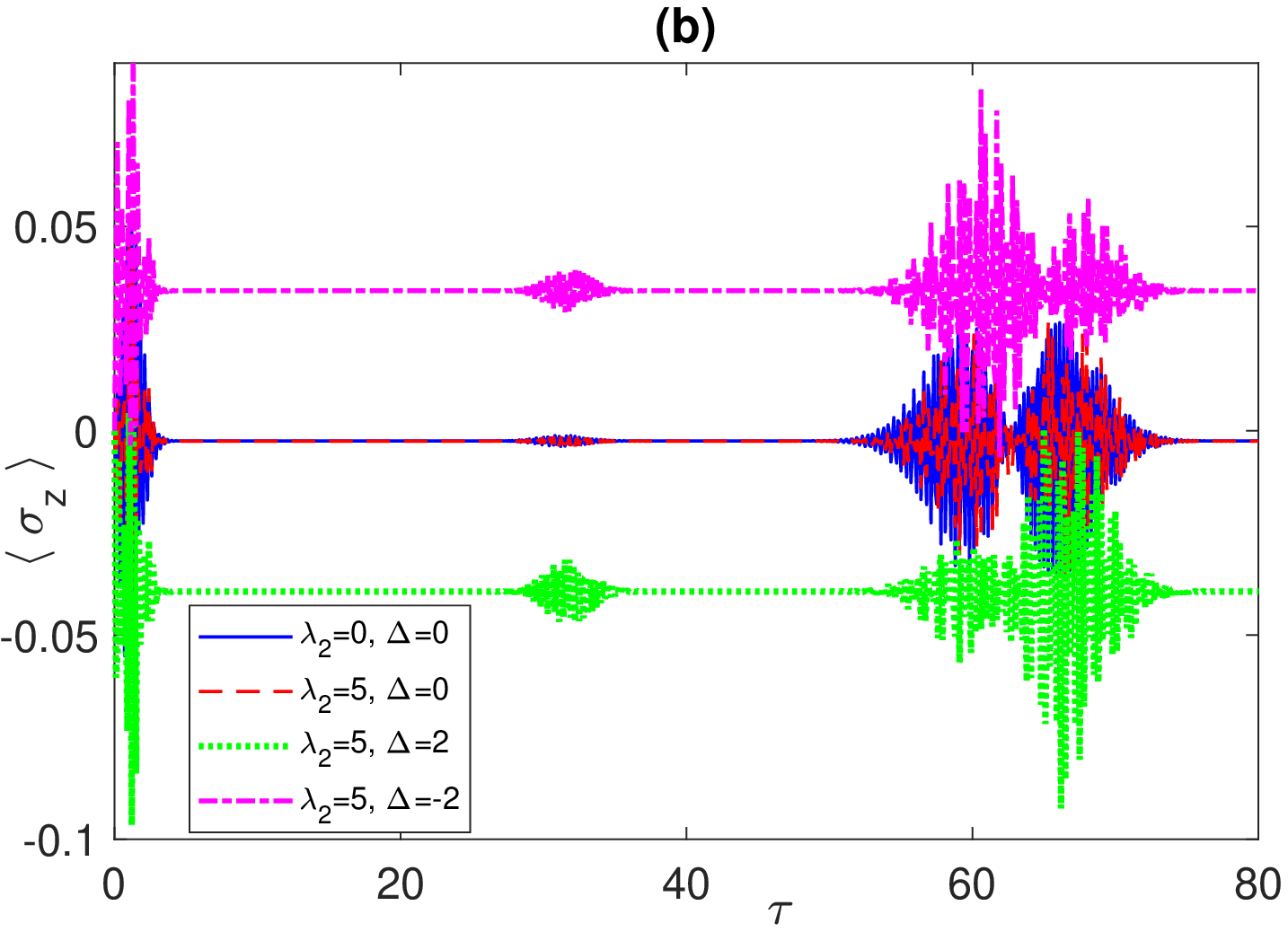}\end{subfigure}
\caption{{
Entanglement $E_f$ and population inversion $\langle \sigma_z \rangle$ versus the scaled time $\tau=\lambda_1 t$ with the two atoms are initially  in a correlated Bell state $\psi_{Bc}=(\vert e_{1}\rangle \vert e_{2}\rangle+\vert g_{1}\rangle \vert g_{2}\rangle)/\sqrt{2}$ and the field is in a coherent state:
(a) $E_f$ versus $\tau$ for $\lambda_2=5$, $\bar{n}=100$ and various values of $\Delta$, and
(b) $\langle \sigma_z \rangle$ versus $\tau$ for $\bar{n}=100$ and various values of $\lambda_2$ and $\Delta$.}}
\label{fig3}
\end{figure}

The combined effect of the atom-atom coupling and non-zero detuning on the entanglement and population inversion is considered in Fig.~\ref{fig3}. While, as we have observed in Fig.~\ref{fig2}, the non-zero detuning can neither remove nor reduce the ESD for uncoupled atoms, it may reduce it or even eliminate it completely for coupled atoms. As illustrated in Fig.~\ref{fig3}(a), setting up the detuning to $\Delta=1$ while $\lambda_2=5$, significantly reduces the entanglement death (dashed red line), but increasing $\Delta$ to 2, completely eliminates the entanglement death producing two entanglement peaks (dotted green line). Applying a negative detuning, $\Delta=-2$, yields the same effect of the positive one but with a slightly lower peak (dash-dotted violet line). As can be noticed in Fig.~\ref{fig3}(b), introducing the atom-atom coupling, $\lambda_2=5$ (dash red line) has no noticeable effect on the population dynamics compared with the zero coupling case (blue solid line), except for a quite small shift down in the constant (mean) value. However, introducing a non-zero detuning, $\Delta=2$, leads to a big shift downwards away from the zero value with larger revival amplitude (dotted green line), while a negative detuning, $\Delta=-2$, results in a shift but upwards this time (dash dotted violet line). On the other hand, one can recognize a clear synchronization between the entanglement peaks (shown in the inset plot of Fig.~\ref{fig3}(a)) and the population revival oscillations, for the coupled atoms at non-zero detuning, which again indicates that the entanglement revives from death and vanishes again as a result of the exchange of energy between the field and the atoms. However, the reduction or removal of ESD due to atom-atom coupling and non-zero detuning is not accompanied by any atomic population oscillation, which means they induce entanglement, though it is very weak, away from any energy exchange between the atoms and the field.
\begin{figure}[htbp]
\centering
\begin{subfigure}\centering\includegraphics[width=6.6cm]{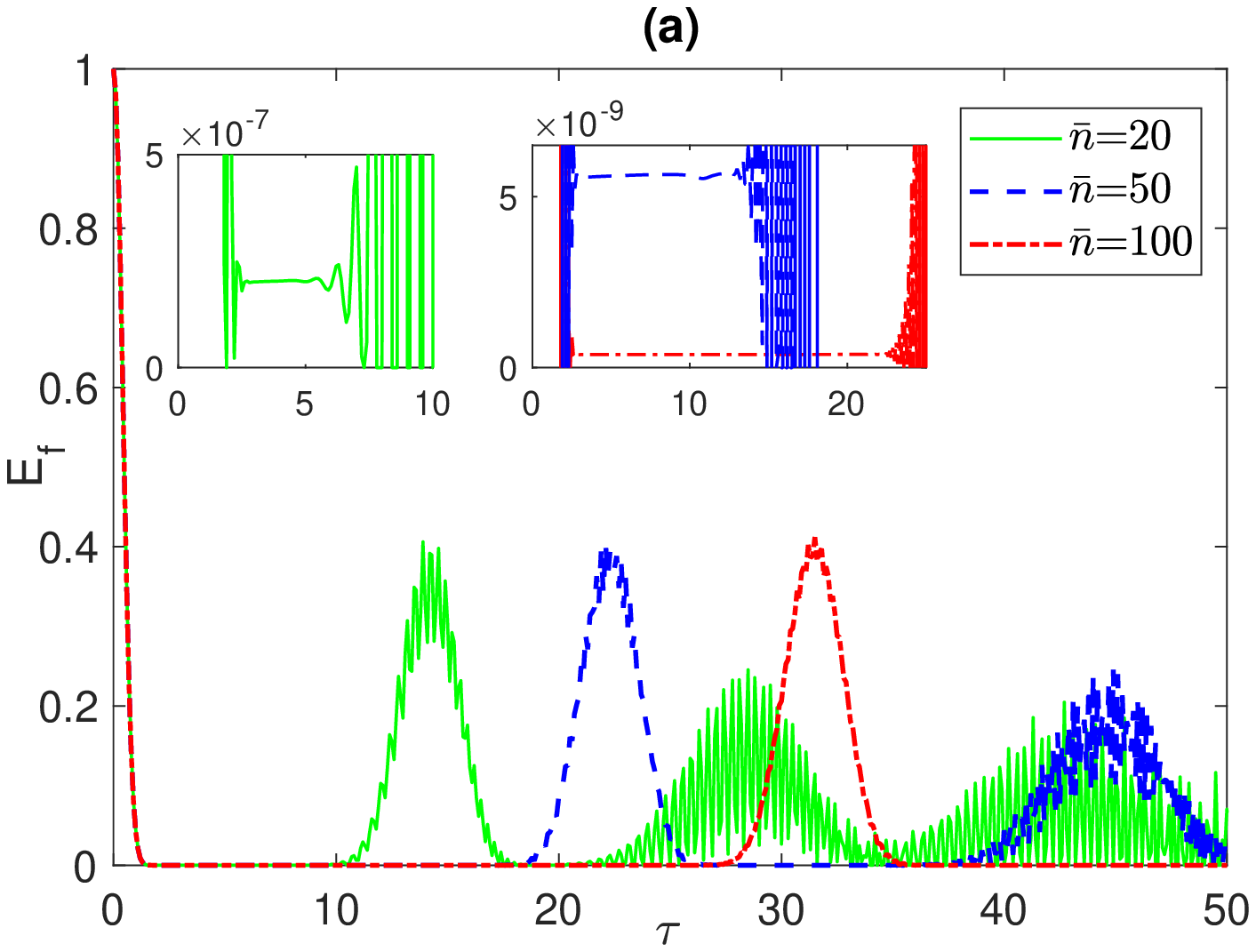}\end{subfigure}
\begin{subfigure}\centering\includegraphics[width=6.6cm]{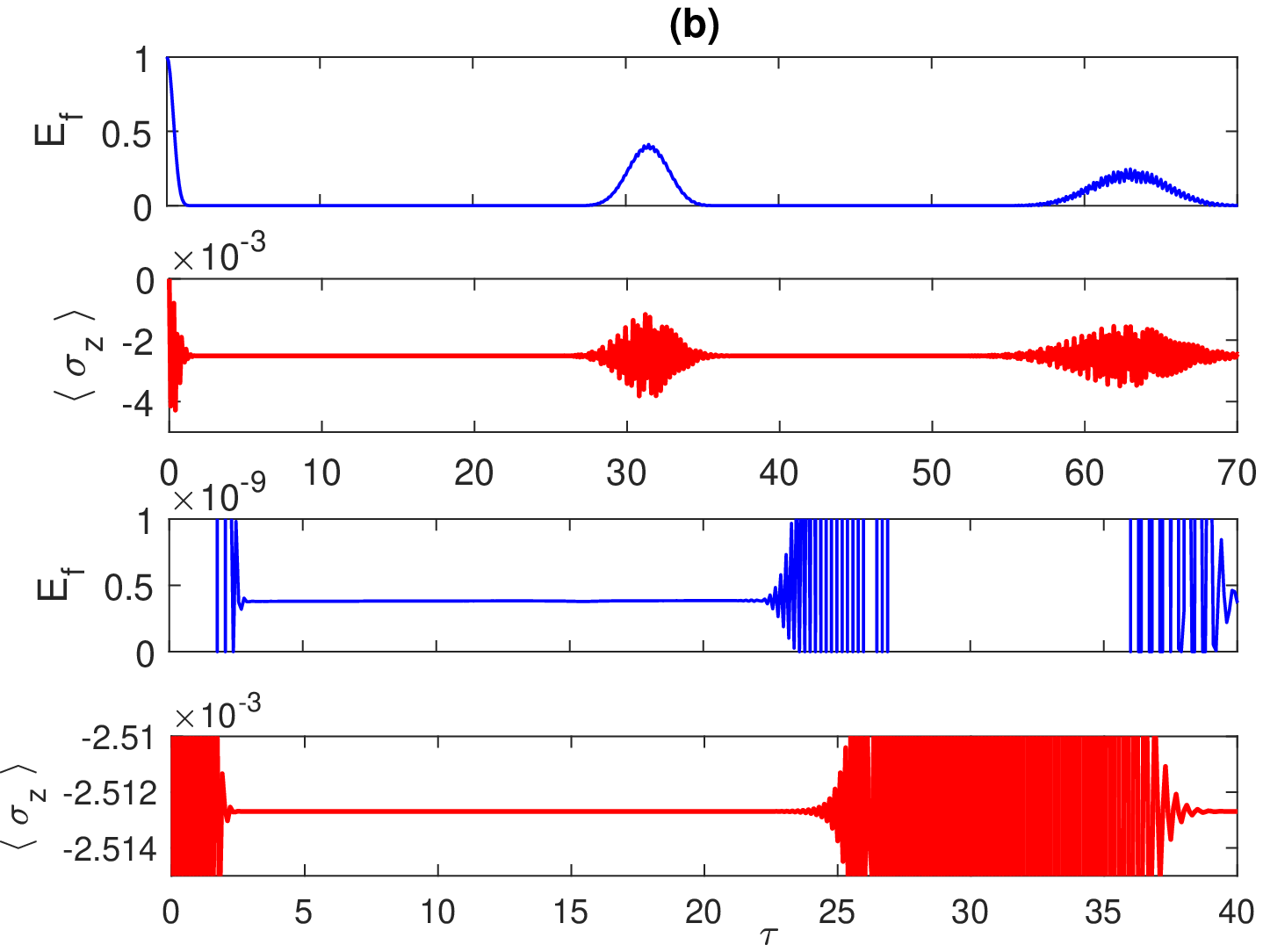}\end{subfigure}
\begin{subfigure}\centering\includegraphics[width=6.6cm]{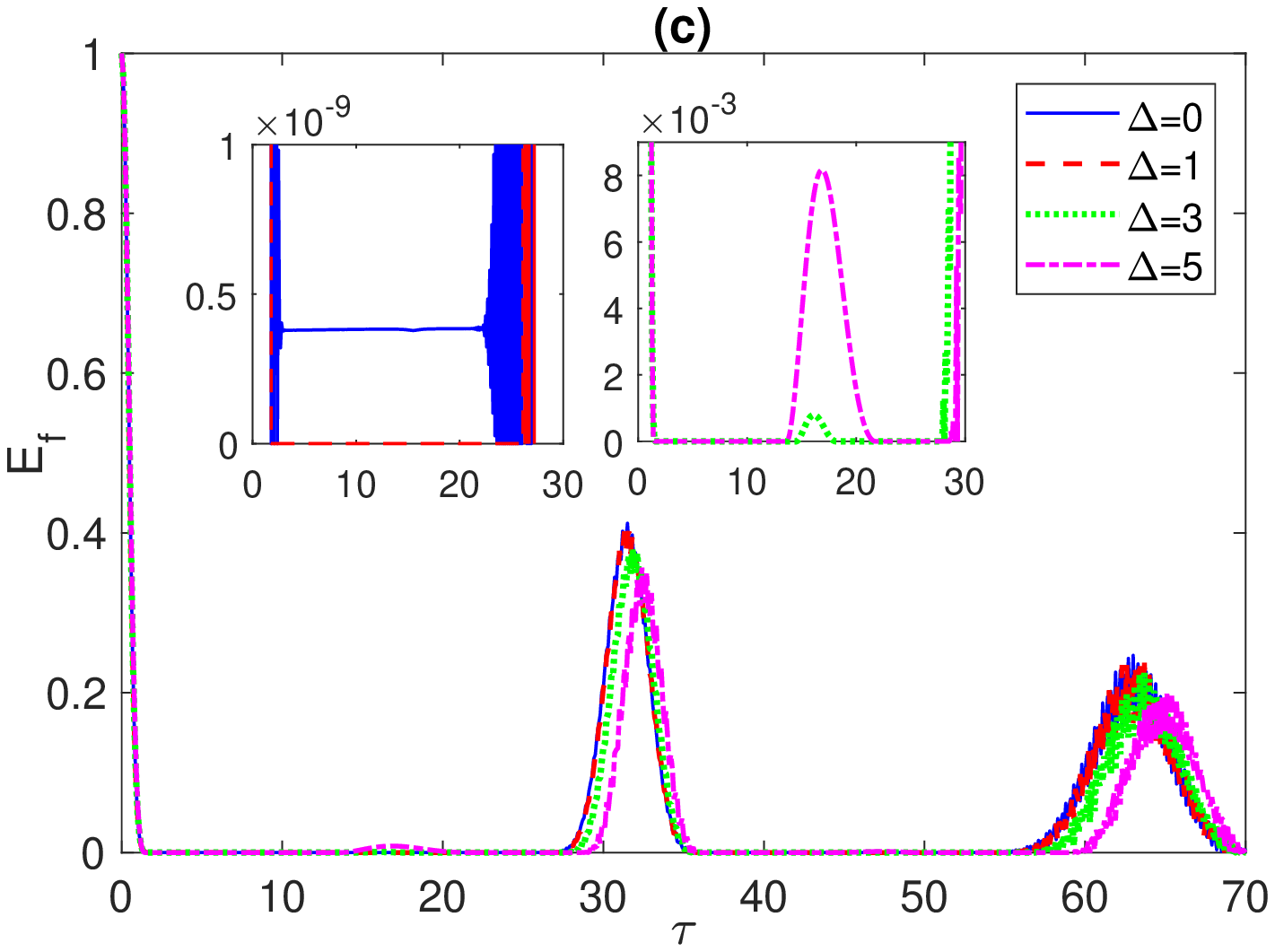}\end{subfigure}
\begin{subfigure}\centering\includegraphics[width=6.6cm]{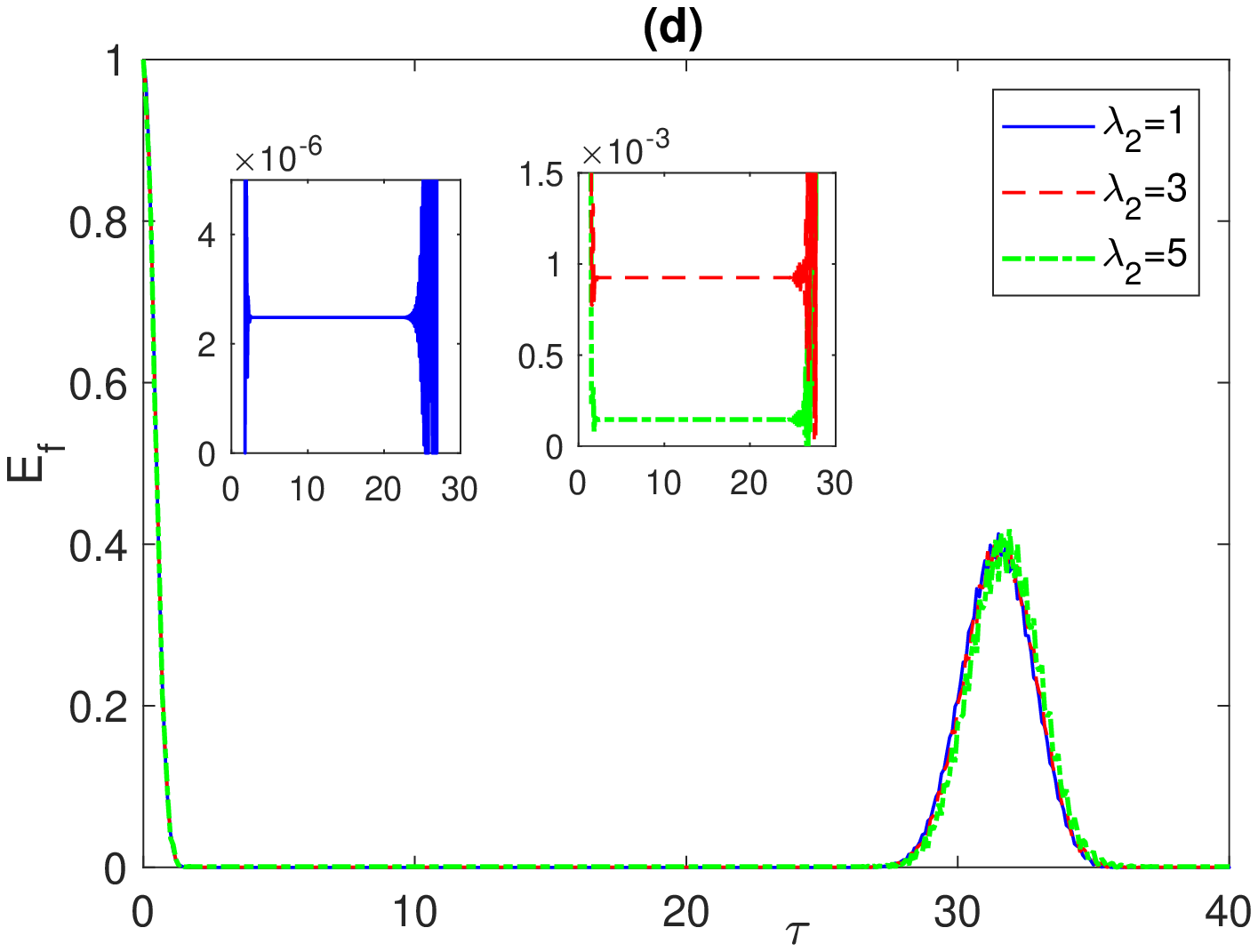}\end{subfigure}
\caption{{
Entanglement $E_f$ and population inversion $\langle \sigma_z \rangle$ versus the scaled time $\tau=\lambda_1 t$ with the two atoms are initially  in an anti-correlated Bell state $\psi_{Ba}=(\vert g_{1}\rangle \vert e_{2}\rangle+\vert e_{1}\rangle \vert g_{2}\rangle)/\sqrt{2}$ and the field is in a coherent state:
(a) $E_f$ versus $\tau$ for $\lambda_2=0$, $\Delta=0$ and various values of the mean number of photons;
(b) $E_f$ and $\langle \sigma_z \rangle$ versus $\tau$ for $\lambda_2=0$, $\Delta=0$ and $\bar{n}=100$; 
(c) $E_f$ versus $\tau$ for $\lambda_2=0$, $\bar{n}=100$ and various values of $\Delta$, and 
(d) $E_f$ versus $\tau$ for $\Delta=0$, $\bar{n}=100$ and various values of $\lambda_2$.}}
\label{fig4}
\end{figure}
In Fig.~\ref{fig4}, we consider a different maximum entanglement initial state that does not yield upon evolution an ESD, for uncoupled atoms at resonance with the field, namely the anti-correlated Bell state $\psi_{Ba}=(\vert g_{1}\rangle \vert e_{2}\rangle+\vert e_{1}\rangle \vert g_{2}\rangle)/\sqrt{2}$. Increasing the radiation field intensity from $\bar{n}=20$ to $50$ and then to $100$ results in very similar impact on the system dynamics to the previous case, the time interval of the very small constant entanglement (rather than zero entanglement in the previous Bell state) increases with radiation intensity as shown in Fig.~\ref{fig4}(a). The inset plots of Fig.~\ref{fig4}(a) illustrates how the constant (mean) entanglement value decreases as the radiation intensity is increased and reaches a value as low as $10^{-9}$ at $\bar{n}=100$. The synchronization between the periods of constant entanglement and the constant population is shown in Fig.~\ref{fig4}(b), which again emphasizes that a steady behavior of the atomic population, where there is no exchange of energy between the atoms and the field leads to a quite small constant entanglement value, while the population revival oscillation boosts the entanglement considerably.
\begin{figure}[htbp]
\centering
\begin{subfigure}\centering\includegraphics[width=6.6cm]{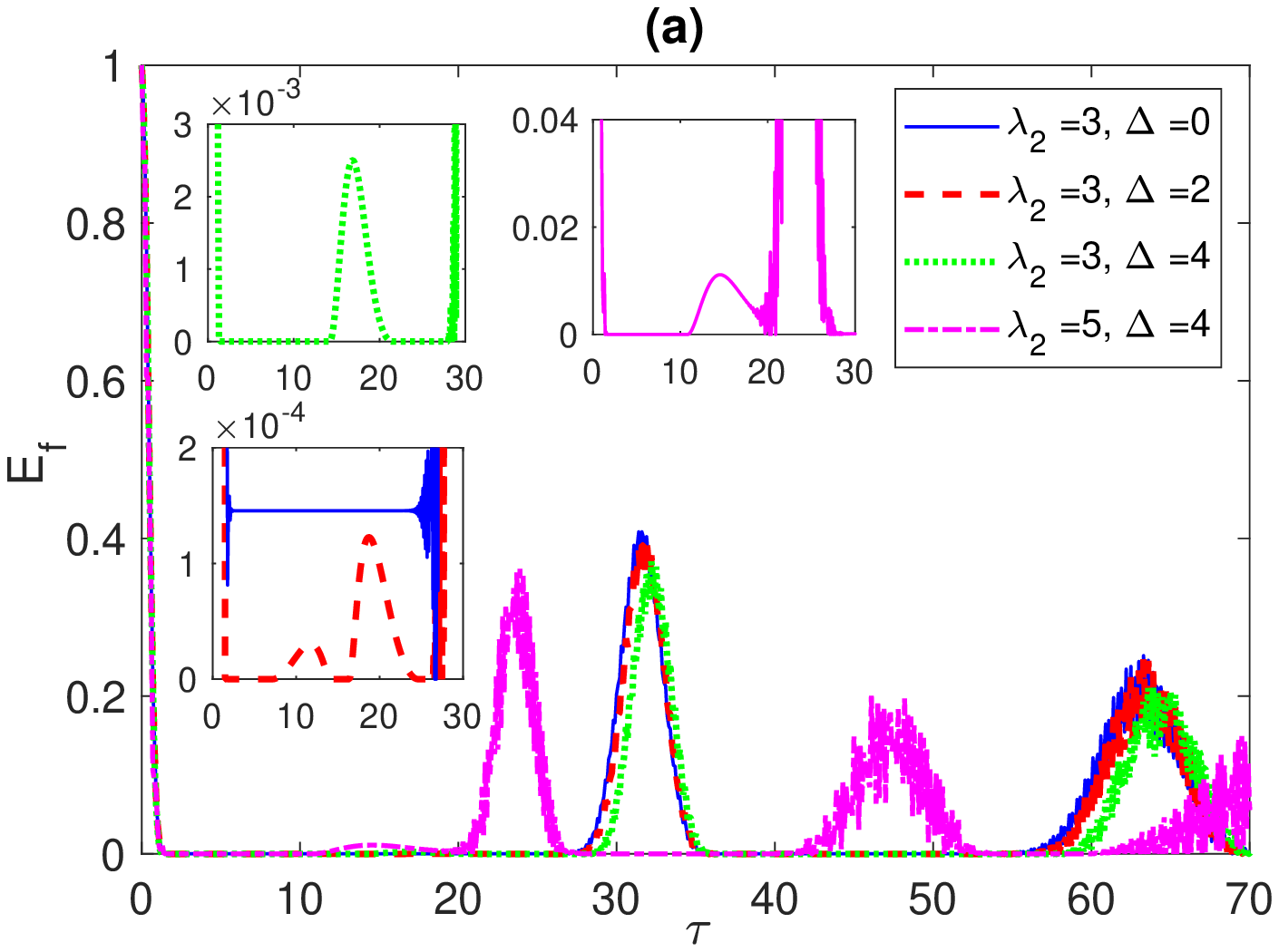}\end{subfigure}
\begin{subfigure}\centering\includegraphics[width=6.6cm]{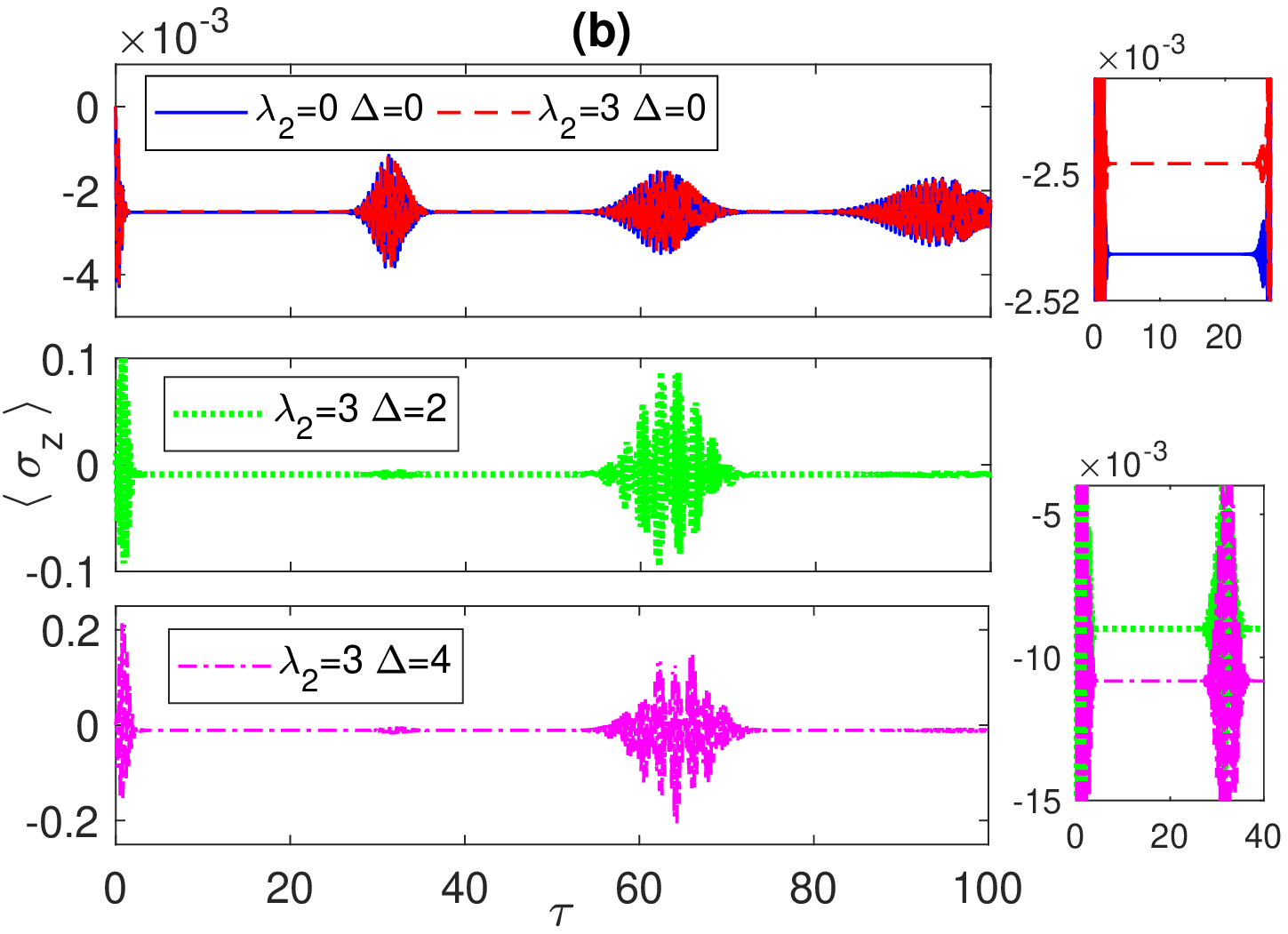}\end{subfigure}
\caption{{
Entanglement $E_f$ and population inversion $\langle \sigma_z \rangle$ versus the scaled time $\tau=\lambda_1 t$ with the two atoms are initially  in an anti-correlated Bell state $\psi_{Ba}=(\vert g_{1}\rangle \vert e_{2}\rangle+\vert e_{1}\rangle \vert g_{2}\rangle)/\sqrt{2}$ and the field is in a coherent state:
(a) $E_f$ versus $\tau$ for $\bar{n}=100$ and various values of $\lambda_2$ and $\Delta$, and 
(b) $\langle \sigma_z \rangle$ versus $\tau$ for $\bar{n}=100$ and various values of $\lambda_2$ and $\Delta$.}}
\label{fig5}
\end{figure}

The effect of non-zero detuning on the entanglement between the uncoupled atoms is depicted in Fig.~\ref{fig4}(c), which shows that a small value of the detuning, $\Delta=1$ (dashed red line) may lead to entanglement death, as shown in the left inset plot, however increasing the detuning further to $\Delta=3$ (dotted green line) induces an intermediate peak within the entanglement death interval which increases considerably when $\Delta$ reaches $5$ (dash dotted violet line), as illustrated in the right inset plot. Switching on the coupling between the two atoms at resonance with the field is considered in Fig.~\ref{fig4}(d), which shows an increase in the constant entanglement value as the coupling is increased from 1 to 3 and finally to 5. Therefore, the atom-atom coupling increases the constant entanglement value while the non-zero detuning leads to entanglement death with intermediate reviving peaks.

The combined effect of $\lambda_2$ and $\Delta$ on the entanglement and the atomic population is illustrated in Fig.~\ref{fig5}(a) and (b) respectively. In the inset plots of Fig.~\ref{fig5}(a), one can see that setting $\lambda_2=3$ and $\Delta=0$ raises the constant entanglement value from the order of $10^{-9}$ (for uncoupled atoms at resonance with the field as shown in Fig.~\ref{fig4}(c)) to another value that is $10^{5}$ higher (solid blue line). Nevertheless, turning on detuning at $\Delta=2$ leads to entanglement death with intermediate reviving peaks (dashed red line), but these peaks turn to a single narrow one with a higher maximum at $\Delta=4$ (dotted green line). However, applying a higher coupling, $\lambda_2=5$,  acts to overcome the detuning effect and partially eliminates the entanglement death while shifting the entanglement reviving peaks to earlier times (dash dotted violet line).
In Fig.~\ref{fig5}(b), we discuss the dynamics of the atomic population, where setting $\lambda_2=3$ at zero detuning (dashed red line) shifts the mean value slightly toward higher value compared with the case of zero coupling (solid blue line), as illustrated in the right top panel. Now turning on a non-zero detuning, $\Delta=2$ at $\lambda_2=3$ shifts the population mean value considerably away from the zero value with a much larger reviving oscillation amplitude (dotted green line). Increasing the detuning parameter to 4, shifts the mean value even further and increases the oscillation amplitude as well (dash dotted violet line), as shown in the right bottom panel. Comparison between the two panels in Fig.~\ref{fig5} emphasizes the synchronization of entanglement dynamics and atomic population.
\begin{figure}[htbp]
\centering
\begin{subfigure}\centering\includegraphics[width=6.6cm]{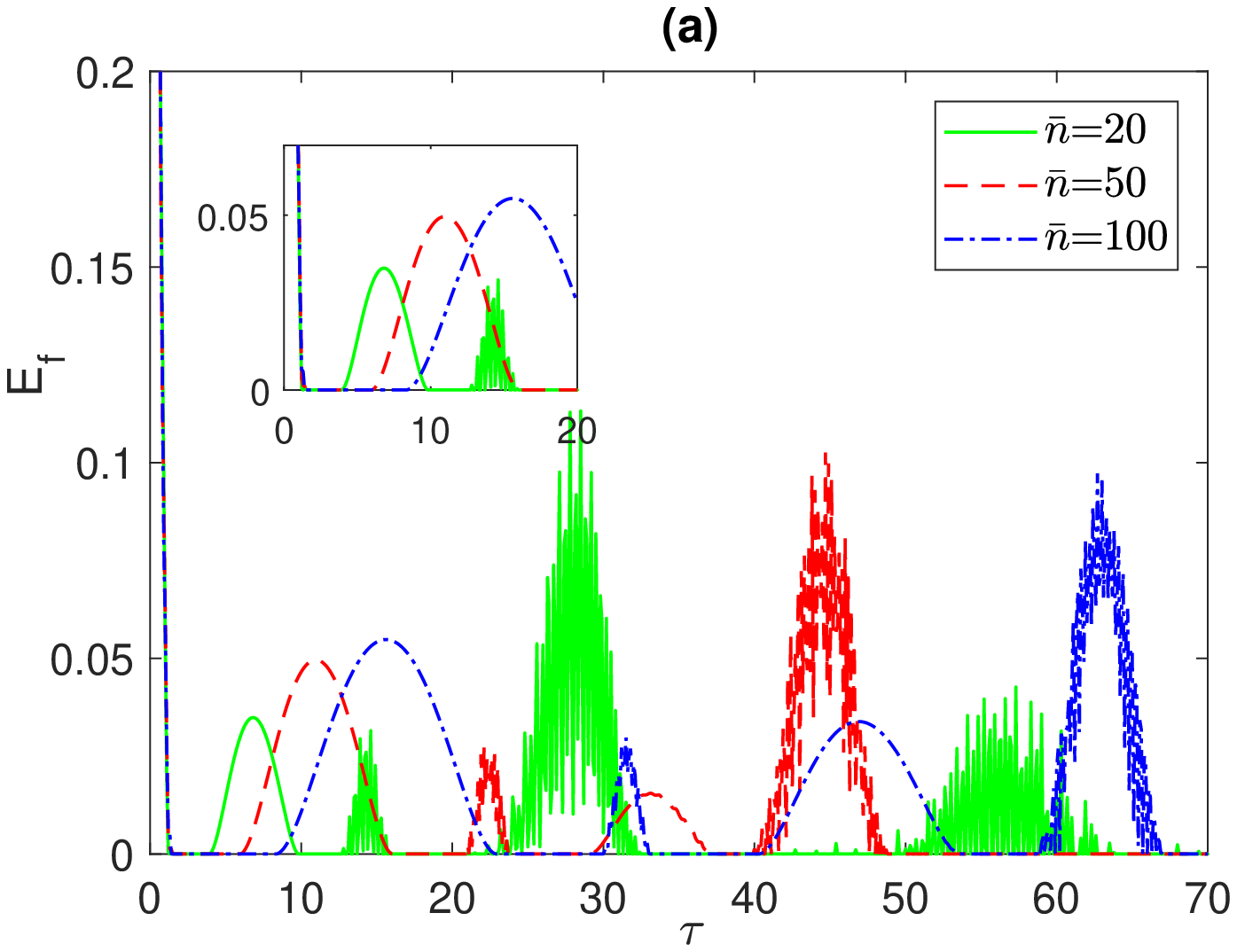}\end{subfigure}
\begin{subfigure}\centering\includegraphics[width=6.6cm]{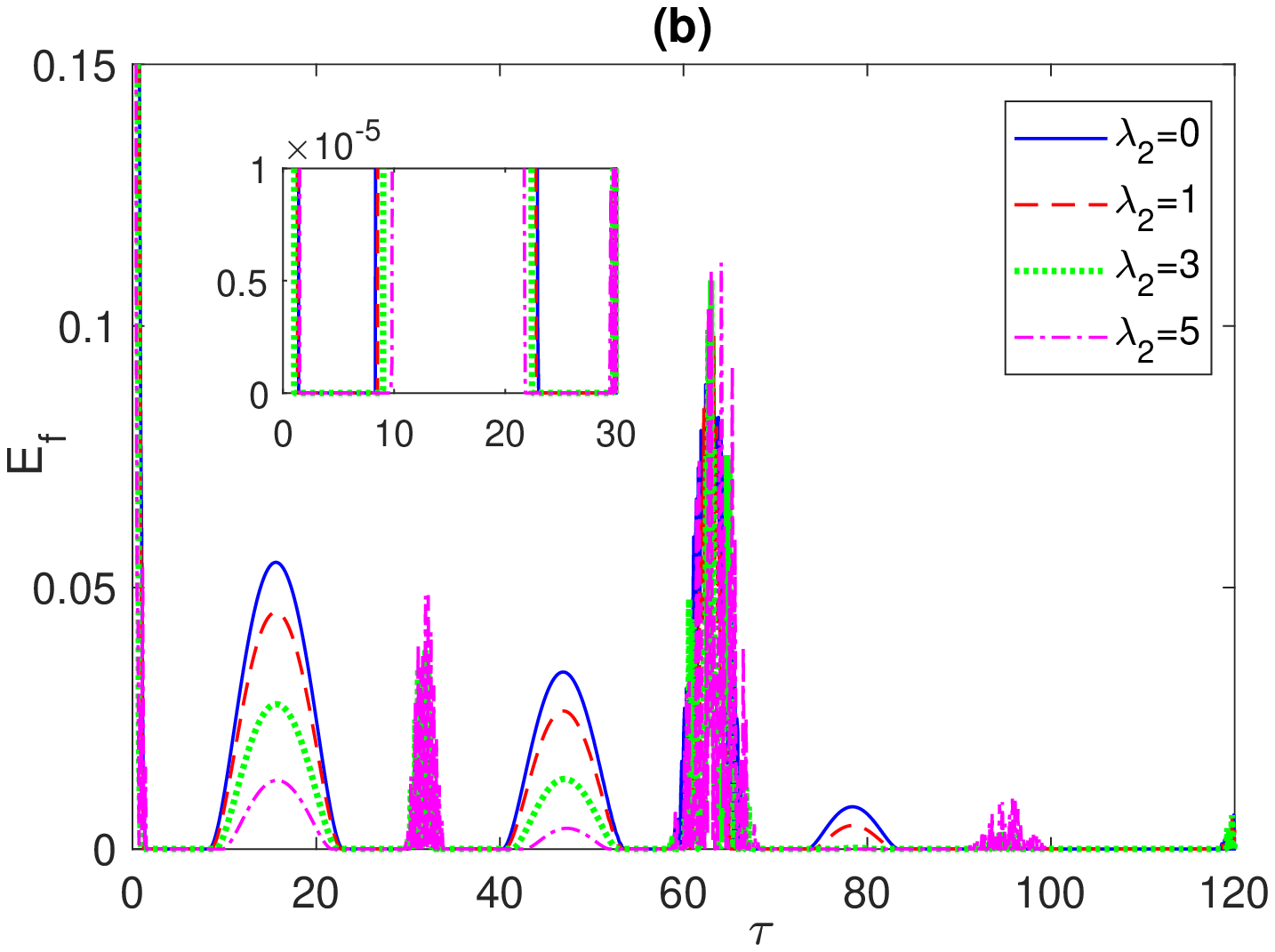}\end{subfigure}
\begin{subfigure}\centering\includegraphics[width=6.6cm]{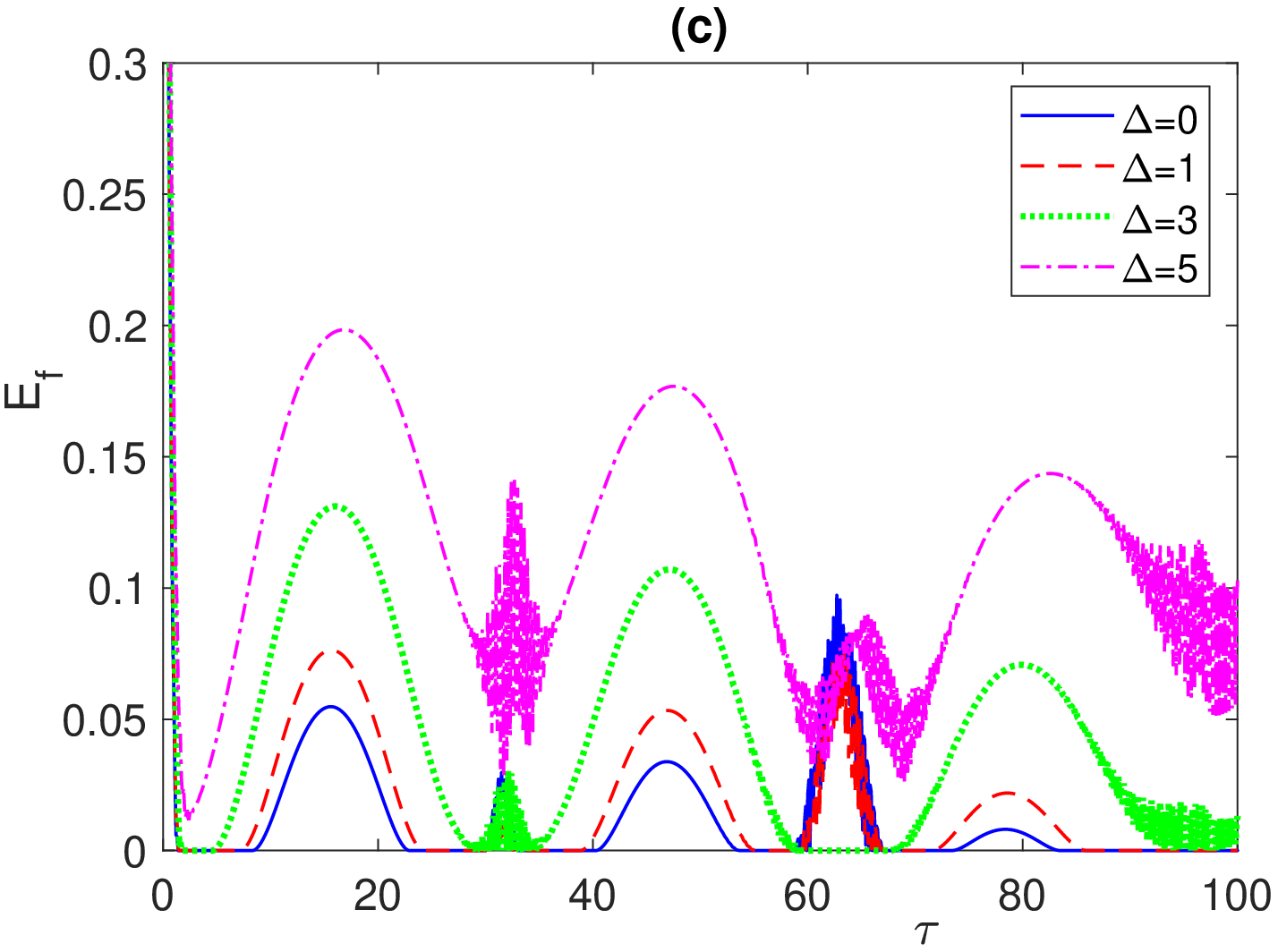}\end{subfigure}
\begin{subfigure}\centering\includegraphics[width=6.6cm]{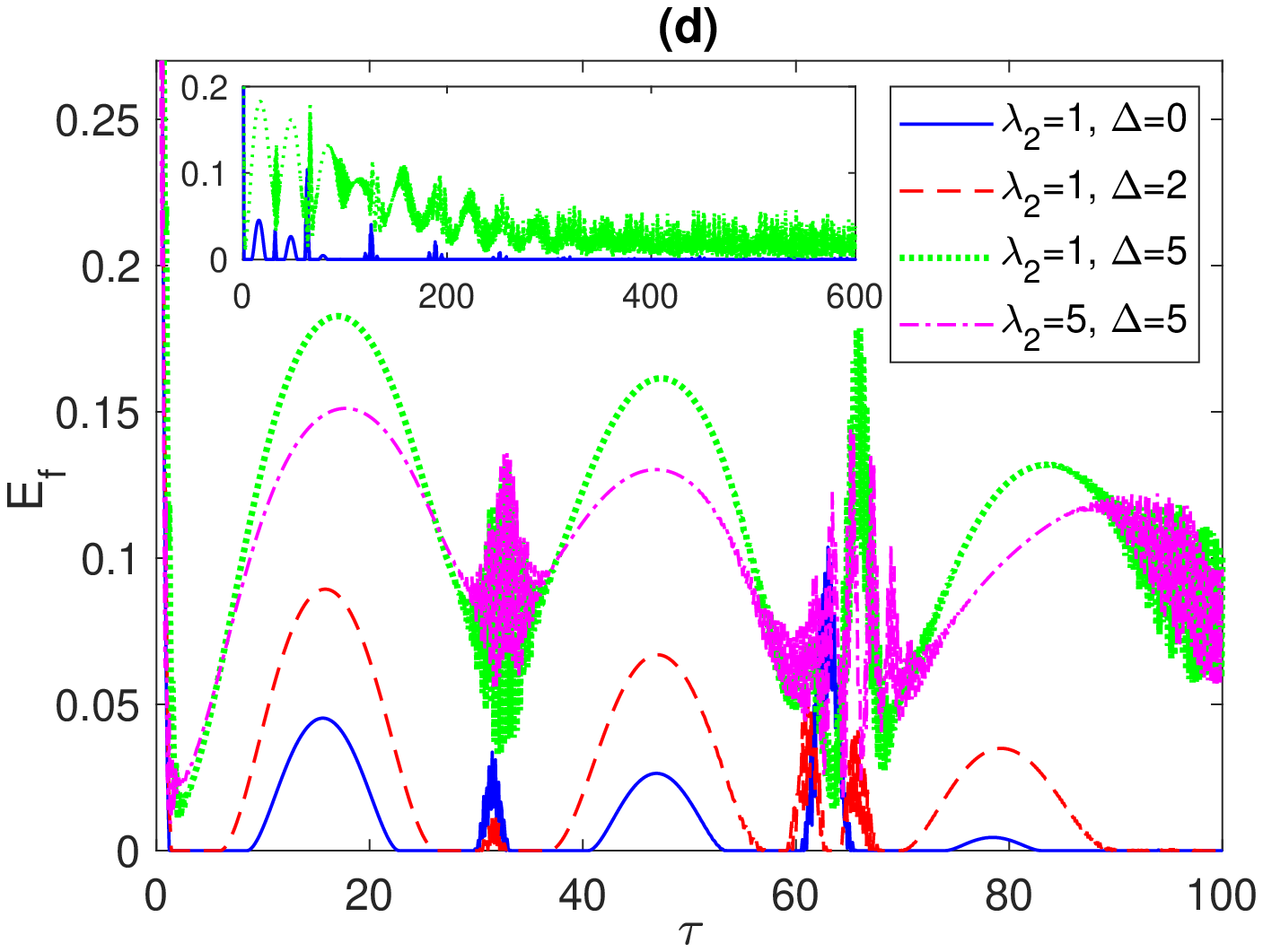}\end{subfigure}
\caption{{
Entanglement $E_f$ versus the scaled time $\tau=\lambda_1 t$ with the two atoms are initially  in a W-like state $\psi_{W}=(\vert g_{1}\rangle \vert g_{2}\rangle + \vert g_{1}\rangle \vert e_{2}\rangle + \vert e_{1}\rangle \vert g_{2}\rangle)/\sqrt{3}$ and the field is in a coherent state for:
(a) $\lambda_2=0$, $\Delta=0$ and various values of the mean number of photons;
(b) $\Delta=0$, $\bar{n}=100$ and various values of $\lambda_2$;
(c) $\lambda_2=0$, $\bar{n}=100$ and various values of $\Delta$, and 
(d) $\bar{n}=100$ and various values of $\lambda_2$ and $\Delta$.}}
\label{fig6}
\end{figure}

\subsection{Partially entangled initial (W) state}
A partially entangled initial state that yields ESD upon evolution, for uncoupled atoms at resonance with the field, is the W-like state $\psi_{W}=(\vert g_{1}\rangle \vert g_{2}\rangle + \vert g_{1}\rangle \vert e_{2}\rangle + \vert e_{1}\rangle \vert g_{2}\rangle)/\sqrt{3}$, which is considered in Fig.~\ref{fig6}. 
In Fig.~\ref{fig6}(a), as can be seen, increasing the radiation field intensity leads to longer ESD time intervals and reduces the entanglement oscillation. Clearly, the ESD time intervals in the current case is much smaller than the ones corresponding to the correlated Bell state. Testing the atom-atom coupling effect on the entanglement dynamics at zero detuning is depicted in Fig.~\ref{fig6}(b), where increasing the coupling strength from 1 to 3 and then to 5, increases the sudden death time interval, which is clearly illustrated in the inset plot, and makes the entanglement reviving peaks narrower with a smaller maximum value, i.e. the atomic coupling enhances ESD. On the other hand, applying a non-zero detuning to uncoupled atoms removes the ESD partially, at $\Delta=3$ (the dotted green line), or even completely, at $\Delta=5$ (dash dotted violet line), as shown in Fig.~\ref{fig6}(c). Therefore, the off-resonance interaction between the field and the atoms can be utilized to completely terminate ESD and the entanglement oscillation in this case indicates a reduction in the atom-atom entanglement (by transfer to the other subsystems) without completely vanishing before being gained back, which is repeated periodically.
\begin{figure}[htbp]
\centering
\begin{subfigure}\centering\includegraphics[width=6.6cm]{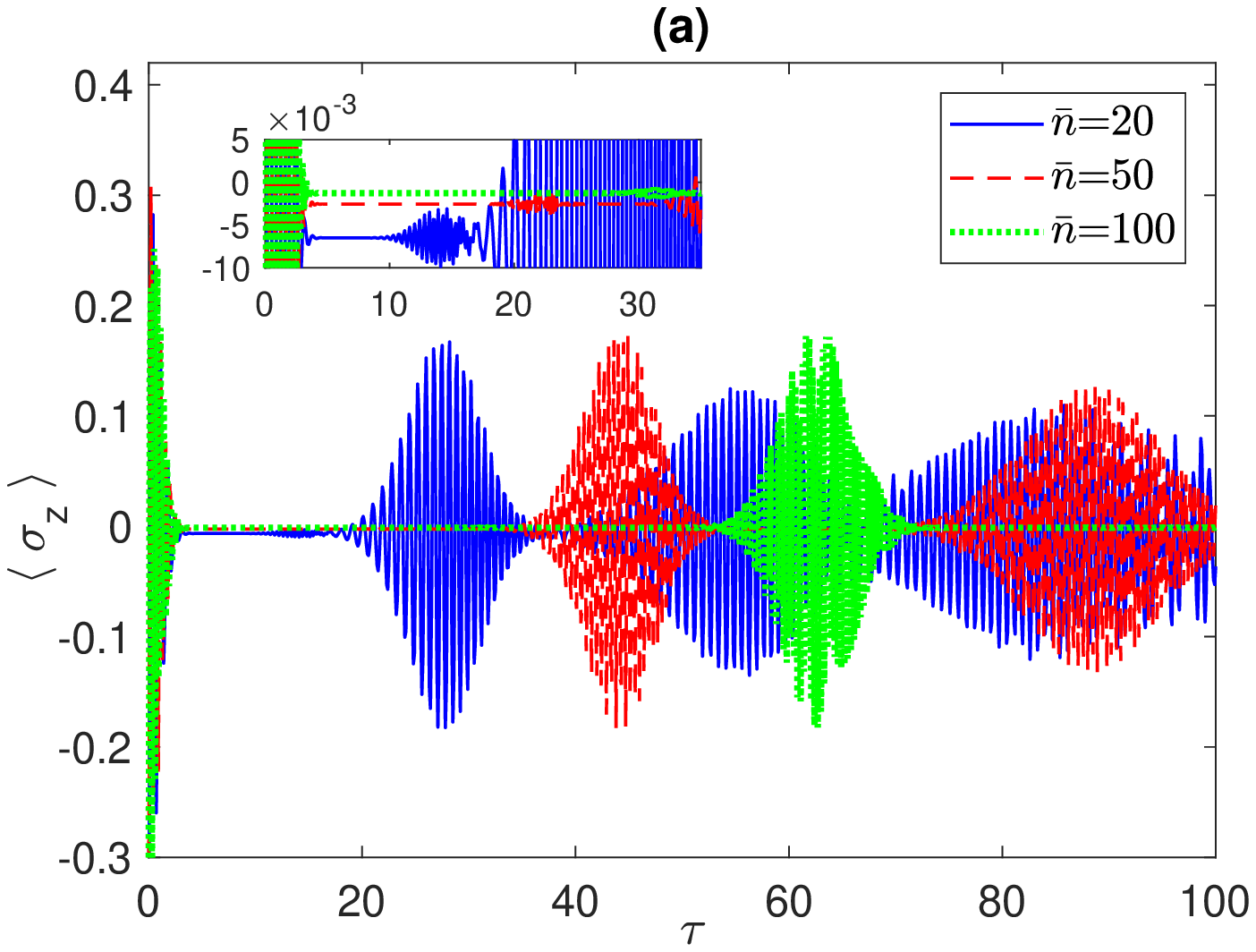}\end{subfigure}
\begin{subfigure}\centering\includegraphics[width=6.6cm]{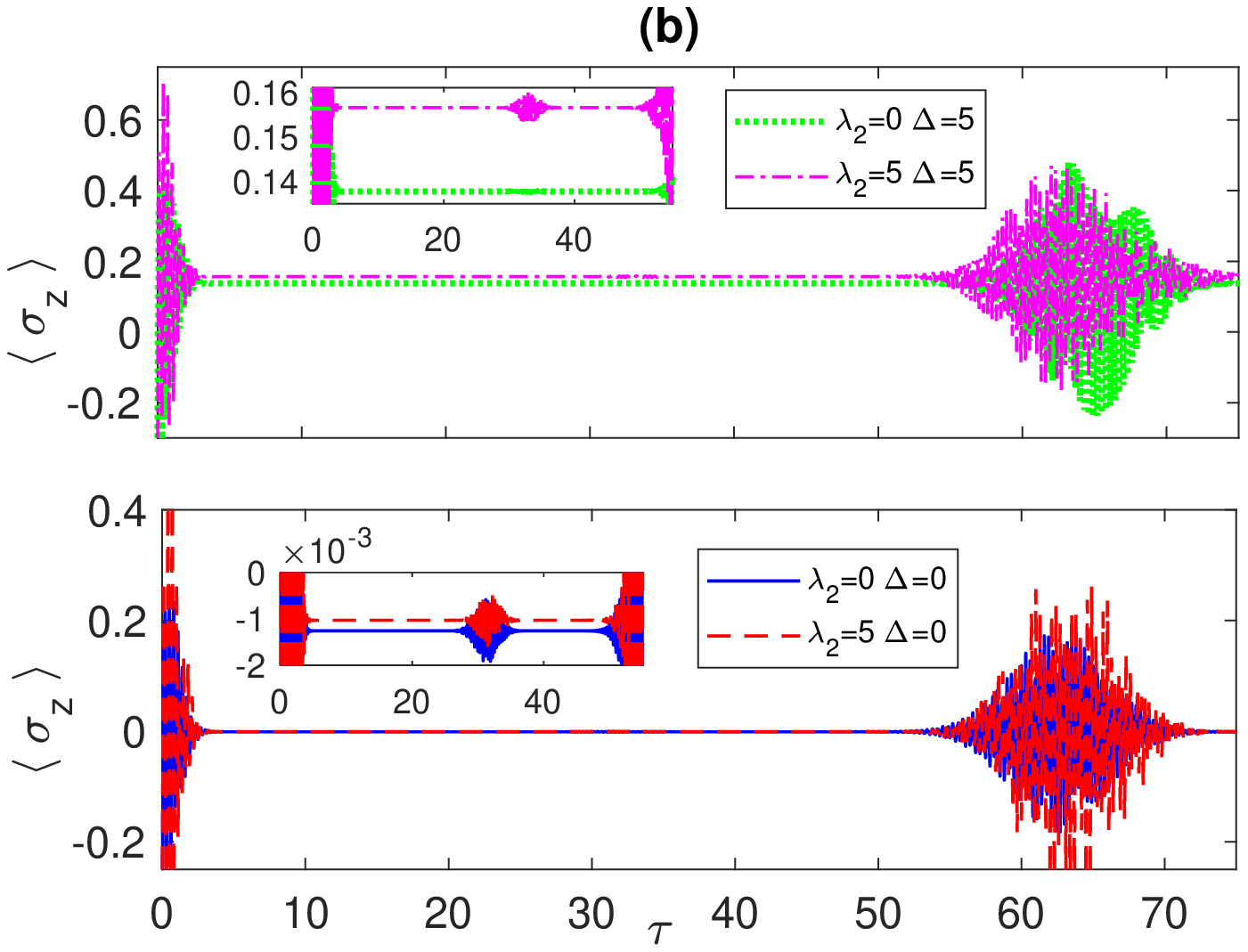}\end{subfigure}
\caption{{
Population inversion versus the scaled time $\tau=\lambda_1 t$ with the two atoms are initially  in a W-like state $\psi_{W}=(\vert g_{1}\rangle \vert g_{2}\rangle + \vert g_{1}\rangle \vert e_{2}\rangle + \vert e_{1}\rangle \vert g_{2}\rangle)/\sqrt{3}$ and the field is in a coherent state for:
(a) $\lambda_2=0$, $\Delta=0$ and various values of the mean number of photons, and
(b) $\bar{n}=100$ and various values of $\lambda_2$ and $\Delta$.}}
\label{fig7}
\end{figure}
The combined presence of atomic coupling and detuning is considered in Fig.~\ref{fig6}(d), in which the atomic coupling suppresses the entanglement, while the non-zero detuning acts the opposite way, eliminating the entanglement death and enhancing the entanglement peaks. As one can notice, setting $\lambda_2=1$ but $\Delta=5$ is enough to completely remove the entanglement death (dotted green line) but increasing the coupling to $\lambda_2=5$  pushes the entanglement peaks down towards the zero value (dash-dotted violet line). The inset plot in Fig.~\ref{fig6}(d) compares the long time behavior of the entanglement for two coupled atoms at resonance with the field, $\lambda_2=1, \Delta=0$ (solid blue line), versus the same two coupled atoms but at non-zero detuning $\Delta=5$, as can be seen the non-zero detuning removes the entanglement death and asymptotically sustains the entanglement oscillation.
\begin{figure}[htbp]
\begin{minipage}[c]{\textwidth}
\centering
\subfigure{\includegraphics[width=10.5cm]{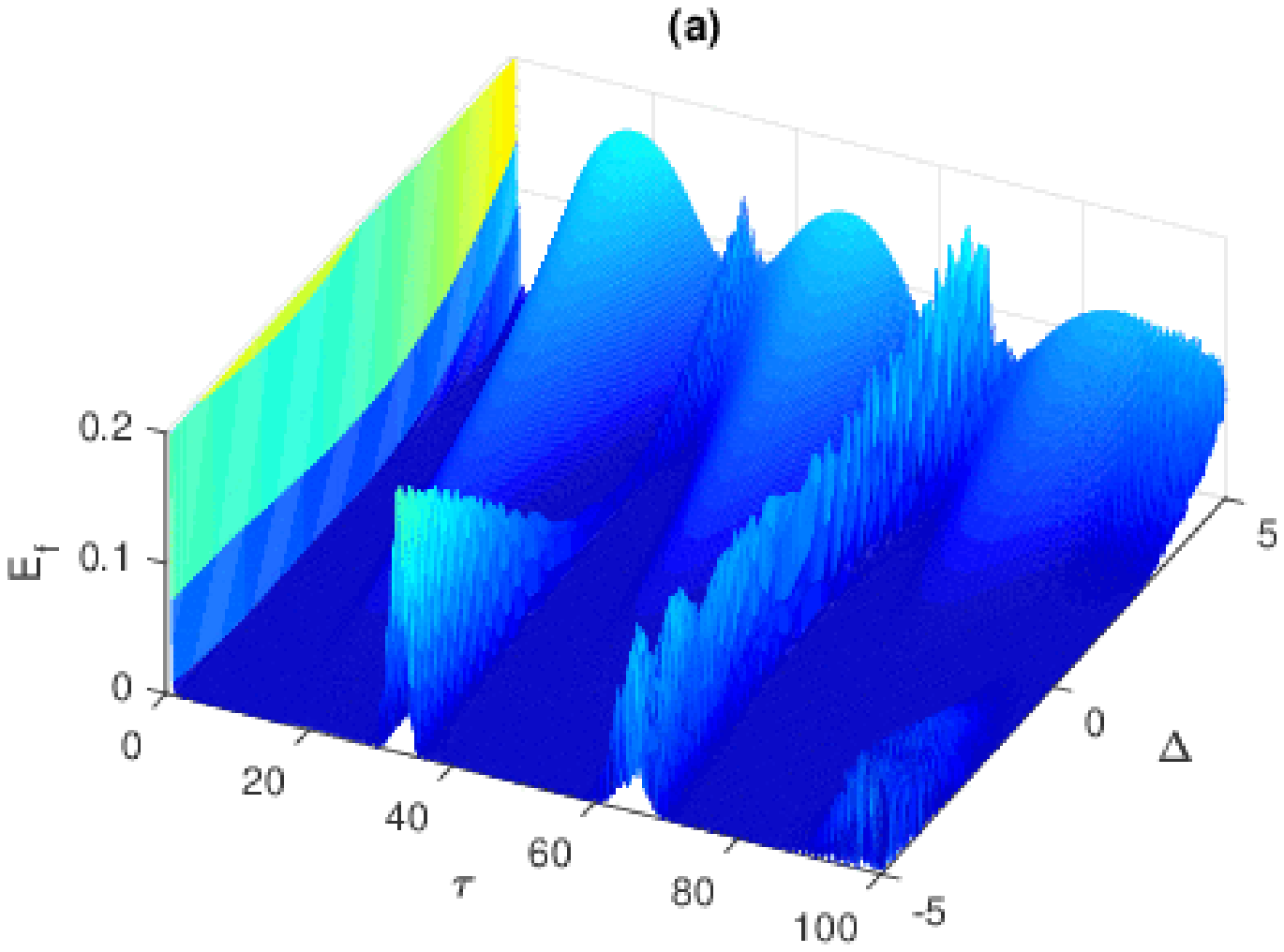}}\quad
\subfigure{\includegraphics[width=10.5cm]{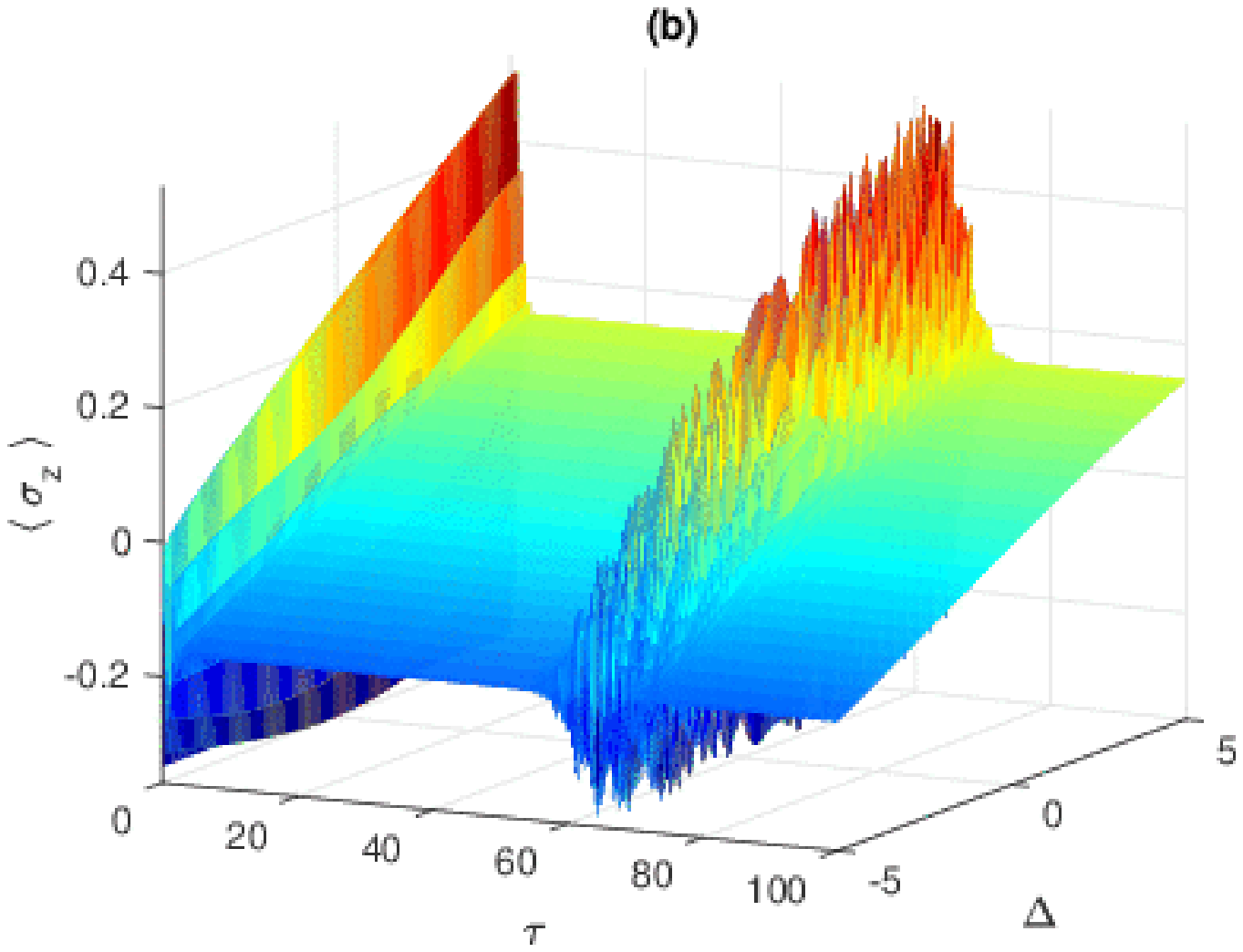}}\quad
\caption{{
Entanglement in (a) and Population inversion in (b) versus the scaled time $\tau=\lambda_1 t$ and the detuning parameter $\Delta$ with the two atoms are initially  in a W-like state $\psi_{W}=(\vert g_{1}\rangle \vert g_{2}\rangle + \vert g_{1}\rangle \vert e_{2}\rangle + \vert e_{1}\rangle \vert g_{2}\rangle)/\sqrt{3}$ and the field is in a coherent state for $\bar{n}=100$ and $\lambda_2=2$.}}
\label{fig8}
\end{minipage}
\end{figure}

In Fig.~\ref{fig7} we discuss the variation in dynamics of the atomic population as a result of changing the system parameters. Increasing the radiation intensity, not only increases the ESD time interval, as illustrated in Fig.~\ref{fig7}(a), but also shifts the population mean value up towards the zero as shown in the inset plot. Figure \ref{fig7}(b) illustrates the effect of atomic coupling, non-zero detuning or both, where as shown in the lower panel and the magnified inset plot, setting $\lambda_2=5$ and $\Delta=0$ produces a very small increment in the mean value of the population towards the zero value and a small increment in the revival oscillation amplitude (dash red line) compared with the uncoupled atoms case (solid blue line). {For uncoupled atoms out of resonance with the field, $\Delta=5$, the mean value of the population is displaced considerably above the zero value with a slight increase in the collapse interval (dotted green line), as illustrated in the upper panel and its inset plot. In the same panel, we consider both of atomic coupling and non-zero detuning, $\lambda_2=5$ and $\Delta=5$, the population mean value shifts further up and the early small revival oscillation at around $\tau =35$ is enhanced (dash dotted violet line).
Comparing the dynamics of entanglement versus atomic population using Figs.~\ref{fig6} and \ref{fig7} (particularly the inset plots of Figs.~\ref{fig7}(b)), one can see that, at either zero or small $\Delta$, ESD is present but its interval is interrupted with a smooth entanglement peak without a corresponding population oscillation. However, when an Entanglement oscillating revival peak appears latter, it is accompanied by an atomic population revival oscillation. On the other hand, when ESD is significantly reduced or even removed, the atomic population collapse does not correspond to either a zero or a quite small entanglement but a peak, while the population revival oscillation correspond to a minimum in the entanglement with a local rapid oscillation, in contrary to what was observed in the correlated Bell state case. Since both of the entanglement and atomic population are very sensitive to the variation in the detuning parameter $\Delta$, as can be noticed in the last two figures, it would be interesting to monitor their dynamic profile over a wide range of (negative and positive) values of $\Delta$, which is illustrated in Fig.~\ref{fig8}. As can be noticed, in Fig.~\ref{fig8}(a), only positive values of $\Delta$ can eliminate the ESD and enhance entanglement as $\Delta$ increases. On the other hand, the atomic population reaches negative values for negative detuning and positive for positive ones, while the amplitude of the revival oscillation increases as $\Delta$ increases as shown in Fig.~\ref{fig8}(b).
\subsection{Separable initial states}
Now we turn to another type of initial states, which is completely separable, namely $\psi_{e}=\vert e_{1}\rangle \vert e_{2}\rangle$, where both atoms are originally in their excited state and the field is in a coherent state. The time evolution of entanglement and population inversion, starting form that initial state is depicted in Fig.~\ref{fig9}.
\begin{figure}[htbp]
\centering
\begin{subfigure}\centering\includegraphics[width=6.6cm]{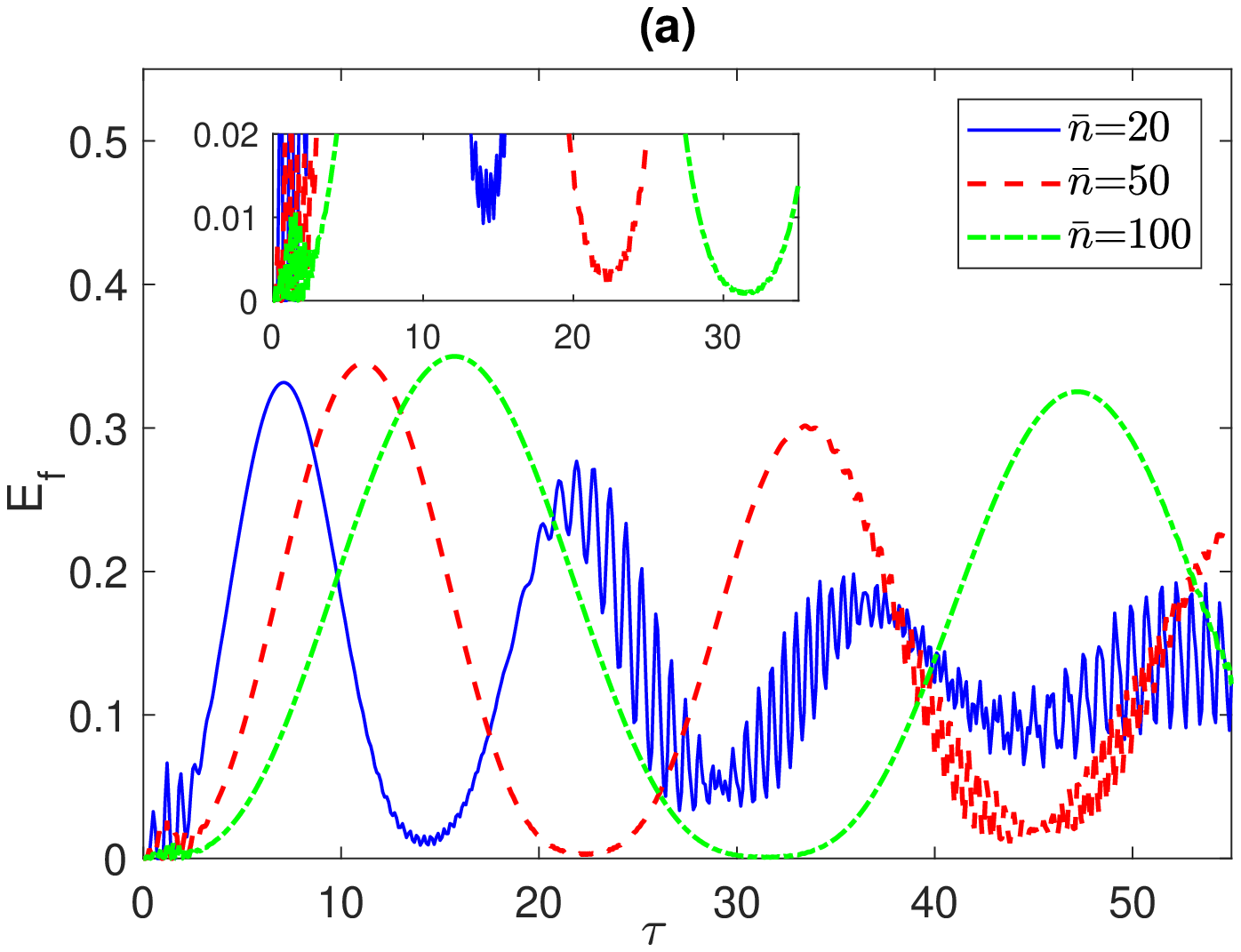}\end{subfigure}
\begin{subfigure}\centering\includegraphics[width=6.6cm]{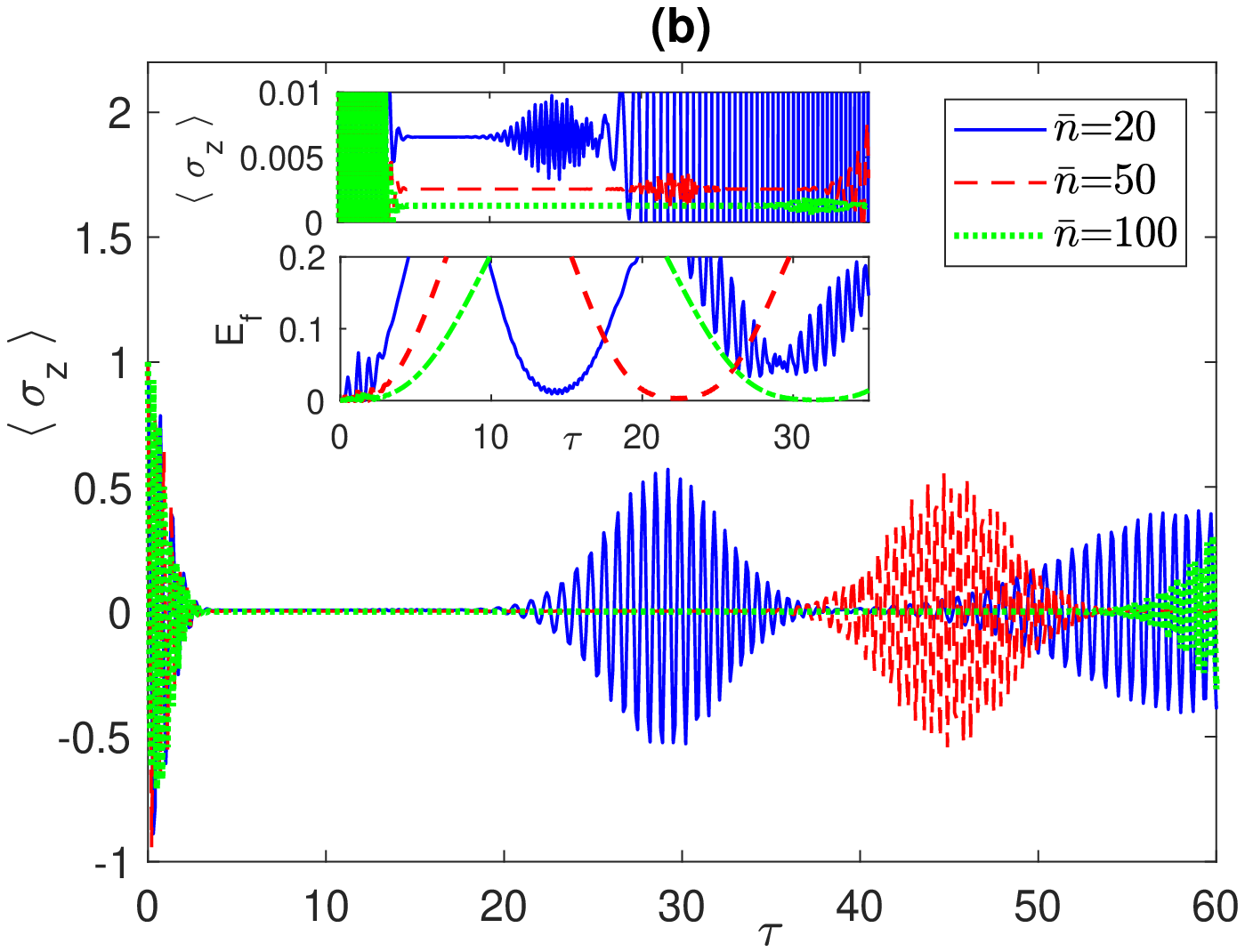}\end{subfigure}
\begin{subfigure}\centering\includegraphics[width=6.6cm]{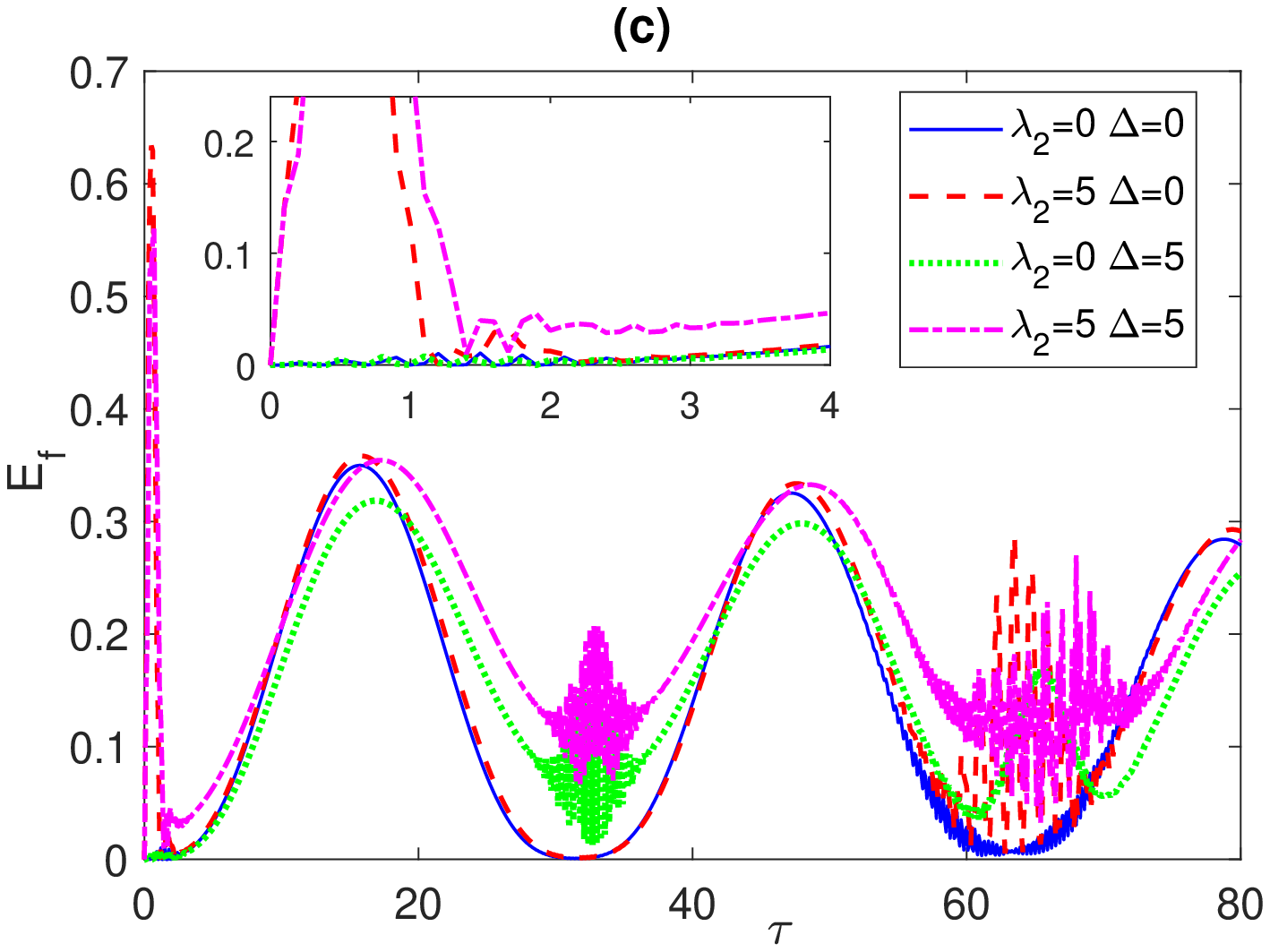}\end{subfigure}
\begin{subfigure}\centering\includegraphics[width=6.6cm]{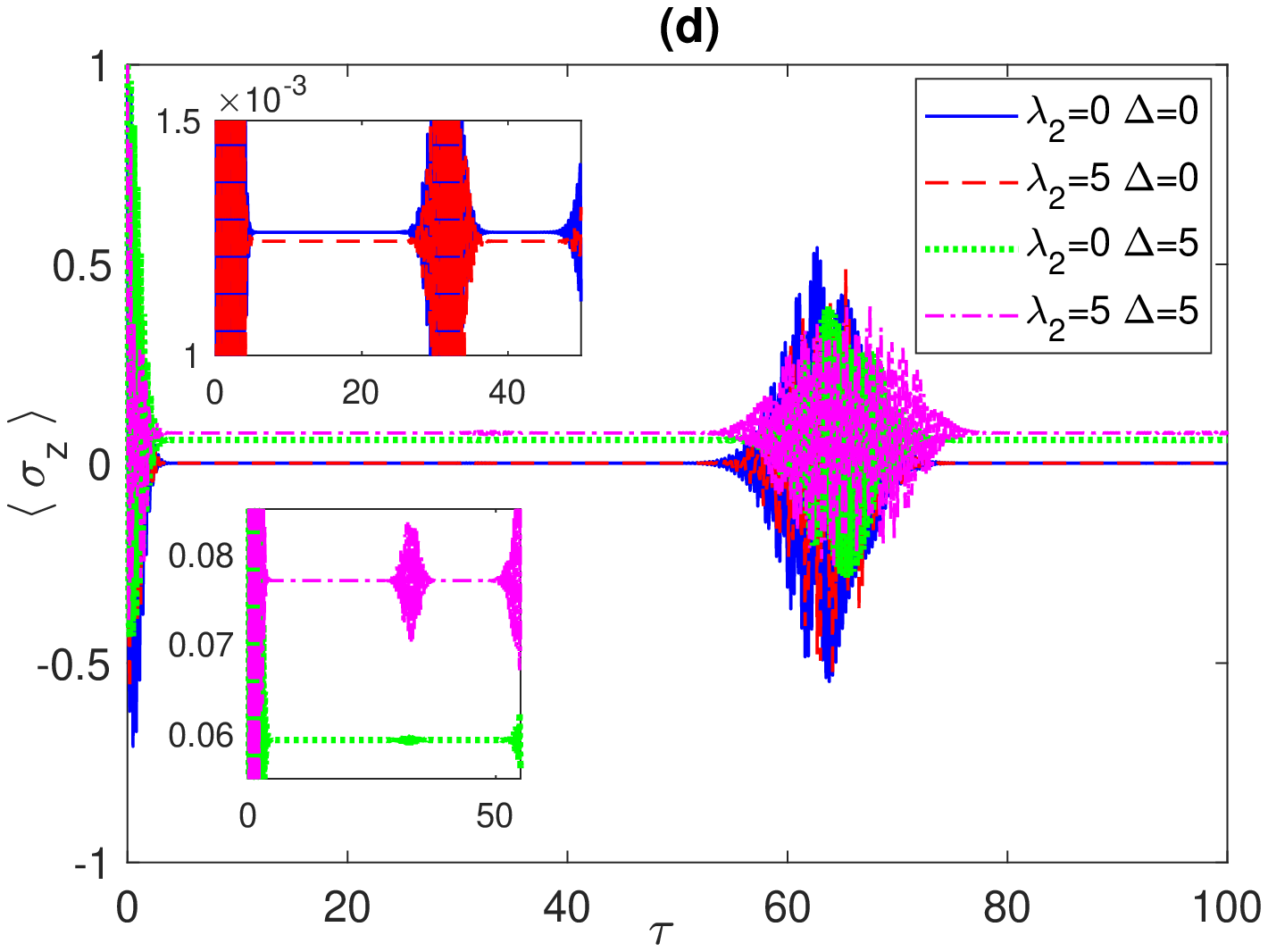}\end{subfigure}
\caption{{
Entanglement $E_f$ and population inversion $\langle \sigma_z \rangle$ versus the scaled time $\tau=\lambda_1 t$ with the two atoms are initially in a disentangled initial state $\psi_{e}=\vert e_{1}\rangle \vert e_{2}\rangle$ and the field is in a coherent state:
(a) $E_f$ versus $\tau$ for $\lambda_2=0$, $\Delta=0$ and various values of the mean number of photons;
(b) $\langle \sigma_z \rangle$ versus $\tau$ for $\lambda_2=0$, $\Delta=0$ and various values of the mean number of photons; 
(c) $E_f$ versus $\tau$ for $\bar{n}=100$ and various values of $\lambda_2$ and $\Delta$, and 
(d) $\langle \sigma_z \rangle$ versus $\tau$ for $\bar{n}=100$ and various values of $\lambda_2$ and $\Delta$.}}
\label{fig9}
\end{figure}
\begin{figure}[htbp]
\begin{minipage}[c]{\textwidth}
\centering
\subfigure{\includegraphics[width=11.5cm]{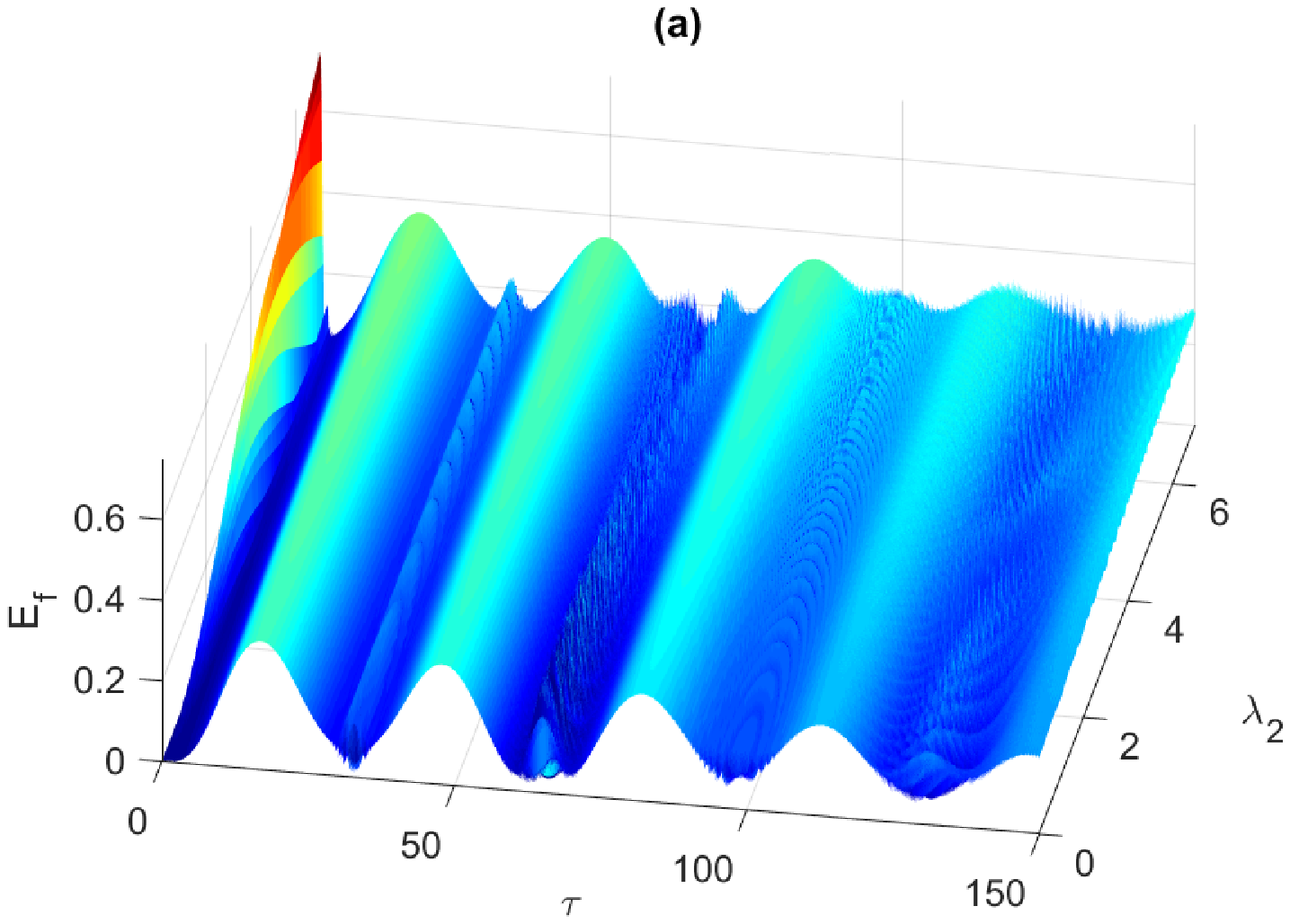}}\quad
\subfigure{\includegraphics[width=11.5cm]{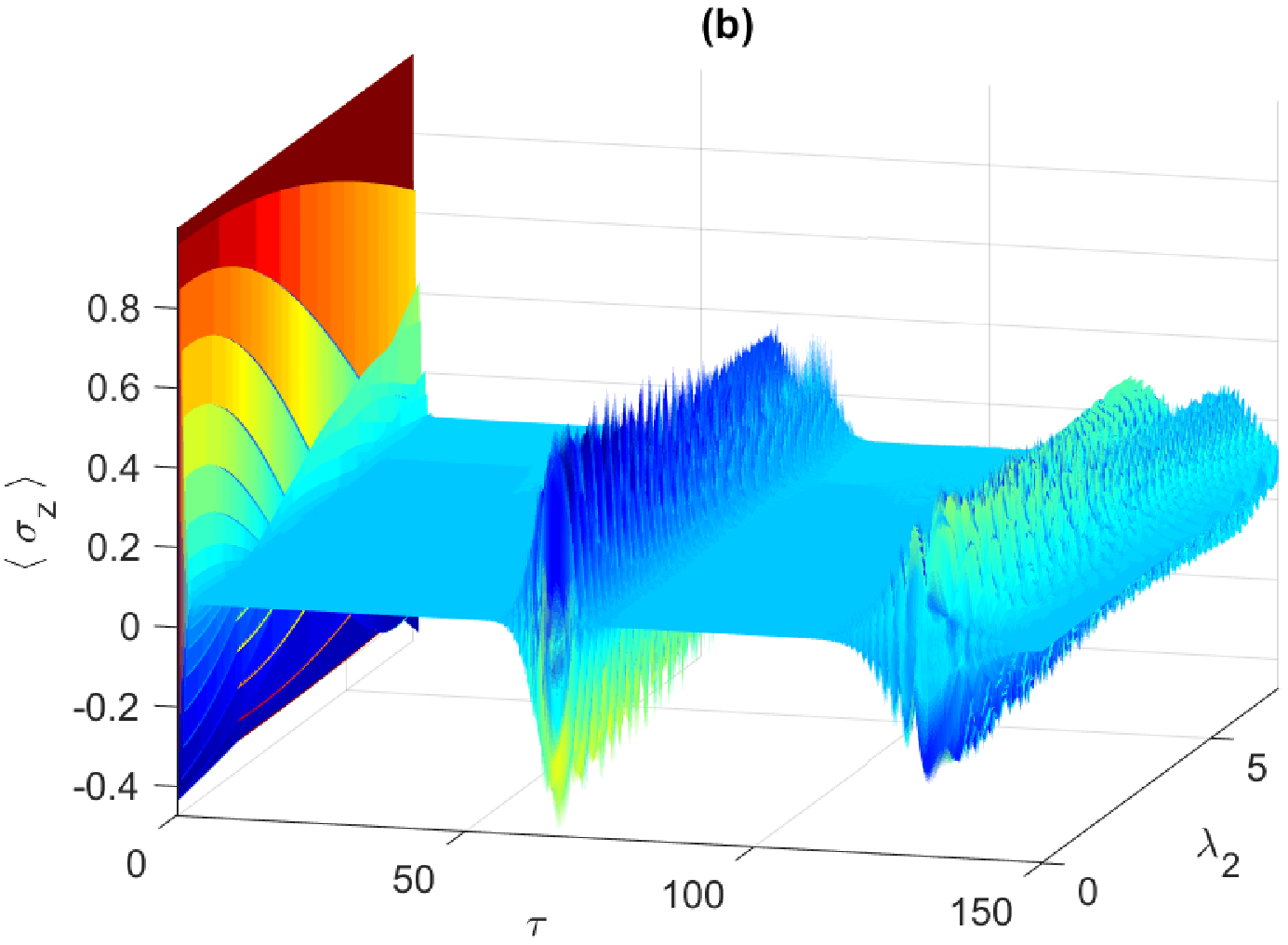}}\quad
\caption{{
Entanglement in (a) and Population inversion in (b) versus the scaled time $\tau=\lambda_1 t$ and the coupling parameter $\lambda_2$ with the two atoms are initially  in a disentangled state $\psi_{e}=\vert e_{1}\rangle \vert e_{2}\rangle$ and the field is in a coherent state for $\bar{n}=100$ and $\Delta=5$.}}
\label{fig10}
\end{minipage}
\end{figure}
For two uncoupled atoms at resonance with the field at $\bar{n}=20$, the entanglement starts from a zero value making a rapid oscillation with a very small amplitude before turning to a larger oscillation with big amplitude then turns to a rapid oscillation again at $\tau \approx 20$ (solid blue line), as illustrated in Fig.~\ref{fig9}(a). But as the intensity of the radiation field is increased, to $\bar{n}=50$ then to 100 (dashed red and dotted green lines respectively), the amplitude of the oscillation slightly increases whereas the frequency decreases to almost its half value. The inset plot of Fig.~\ref{fig9}(a) shows a magnified view of the minima of the entanglement oscillations at different $\bar{n}$ values, which illustrate that none of them reaches zero value. The corresponding change in the population inversion as the field intensity increases is shown in Fig.~\ref{fig9}(b). As can be noticed, higher intensity results in longer collapse time and a slight down shift in the constant collapse value towards the zero. In the inset plots of Fig.~\ref{fig9}(b), we gave a close look and compare the dynamics of $E_f$ and $\langle \sigma_z \rangle$. Clearly there is a strong correlation between the two, where a minimum value of entanglement with a rapid oscillation is corresponding to a revival period of the population inversion, whereas a maximum entanglement corresponds to a collapse period, where is no exchange of energy is taking place between the field and atoms.
\begin{figure}[htbp]
\centering
\begin{subfigure}\centering\includegraphics[width=6.6cm]{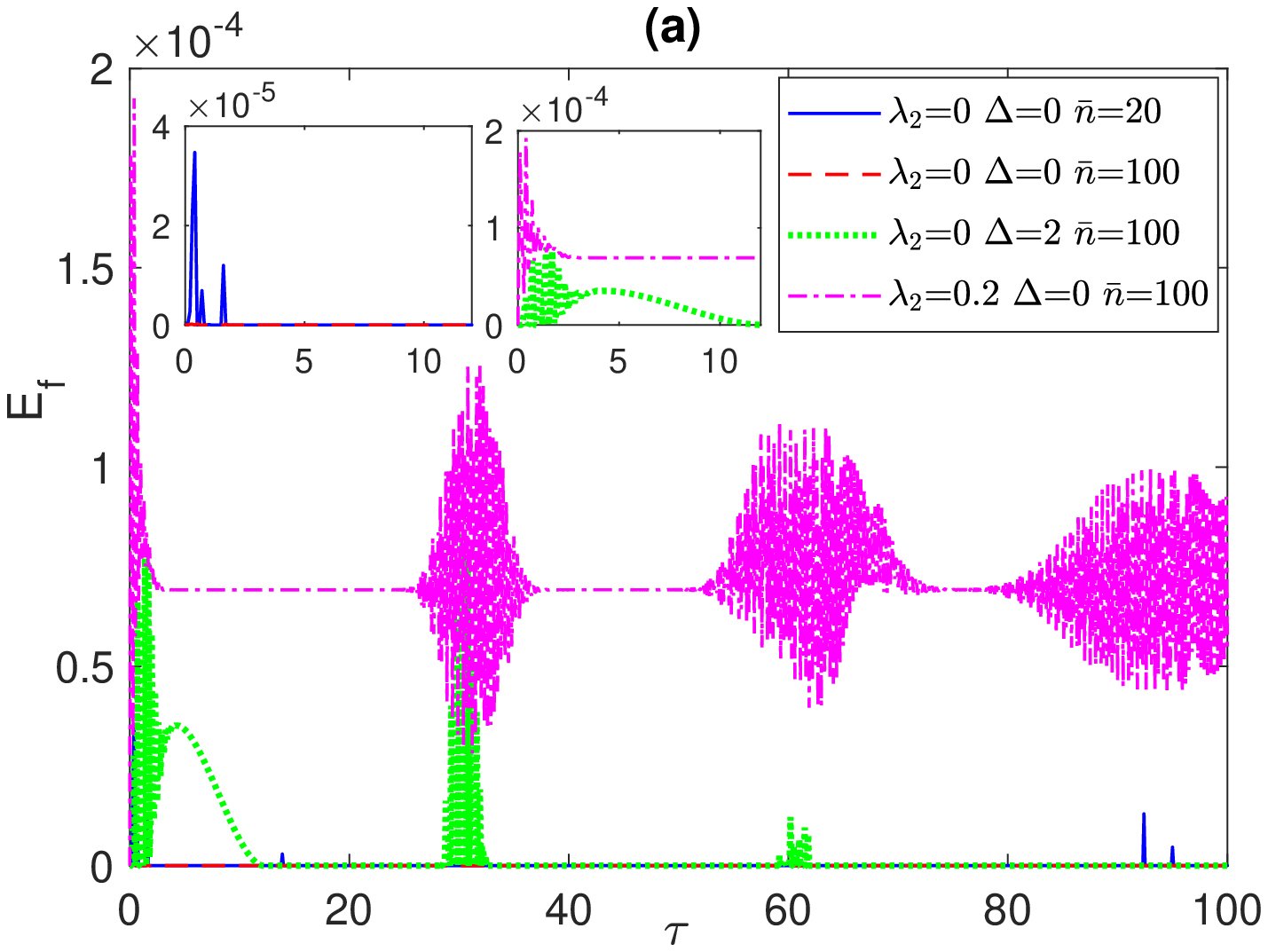}\end{subfigure}
\begin{subfigure}\centering\includegraphics[width=6.6cm]{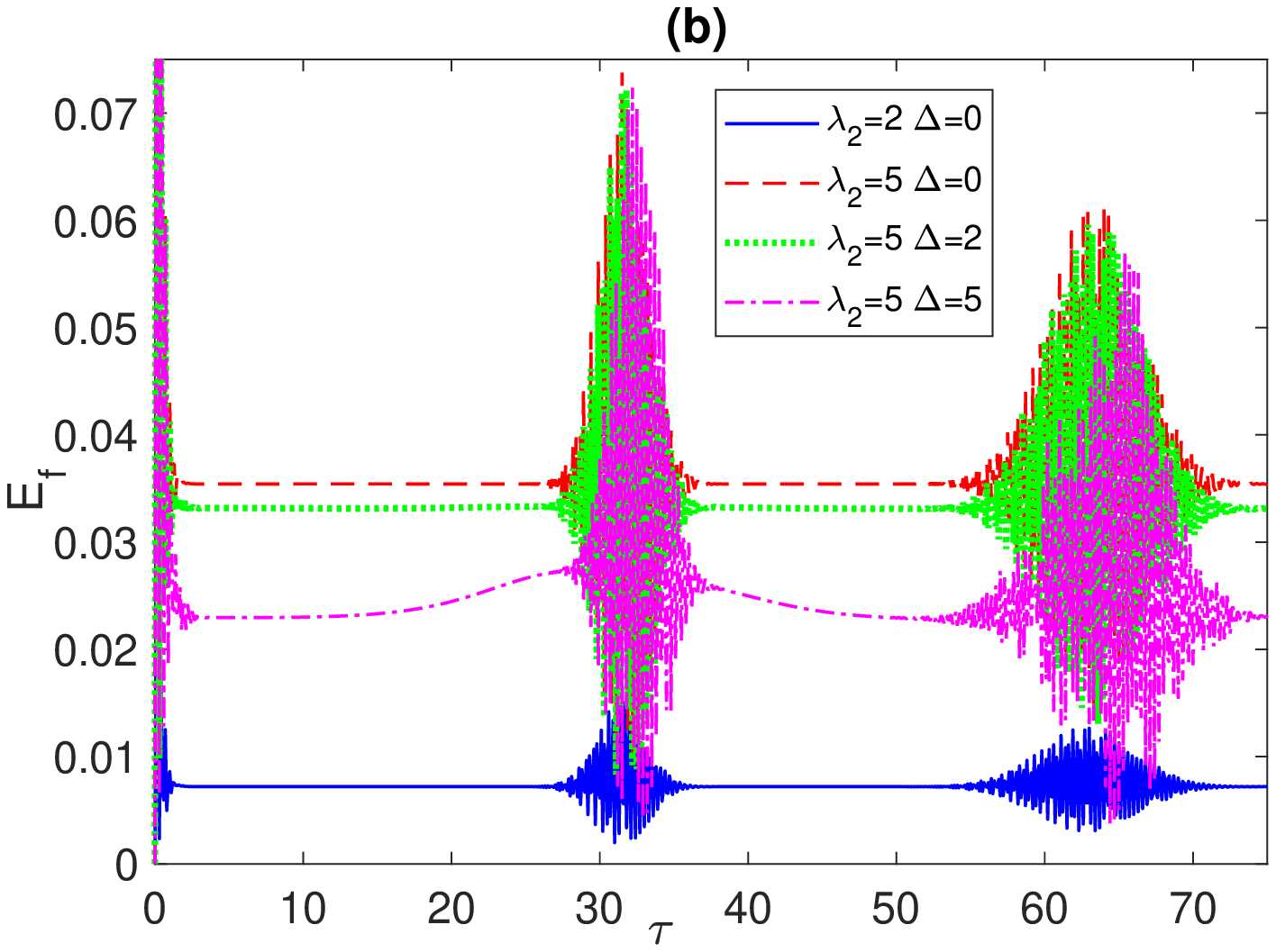}\end{subfigure}
\begin{subfigure}\centering\includegraphics[width=6.6cm]{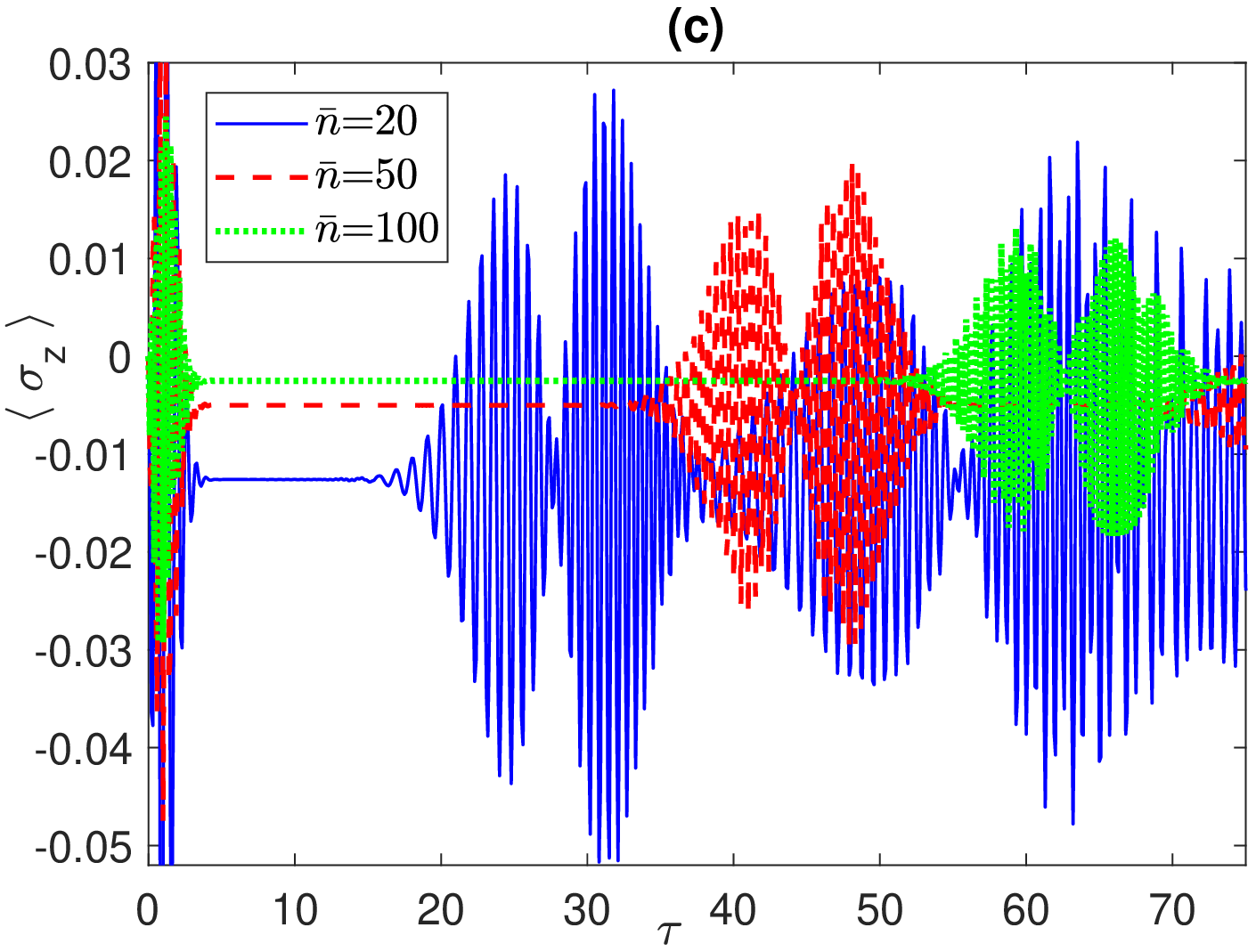}\end{subfigure}
\begin{subfigure}\centering\includegraphics[width=6.6cm]{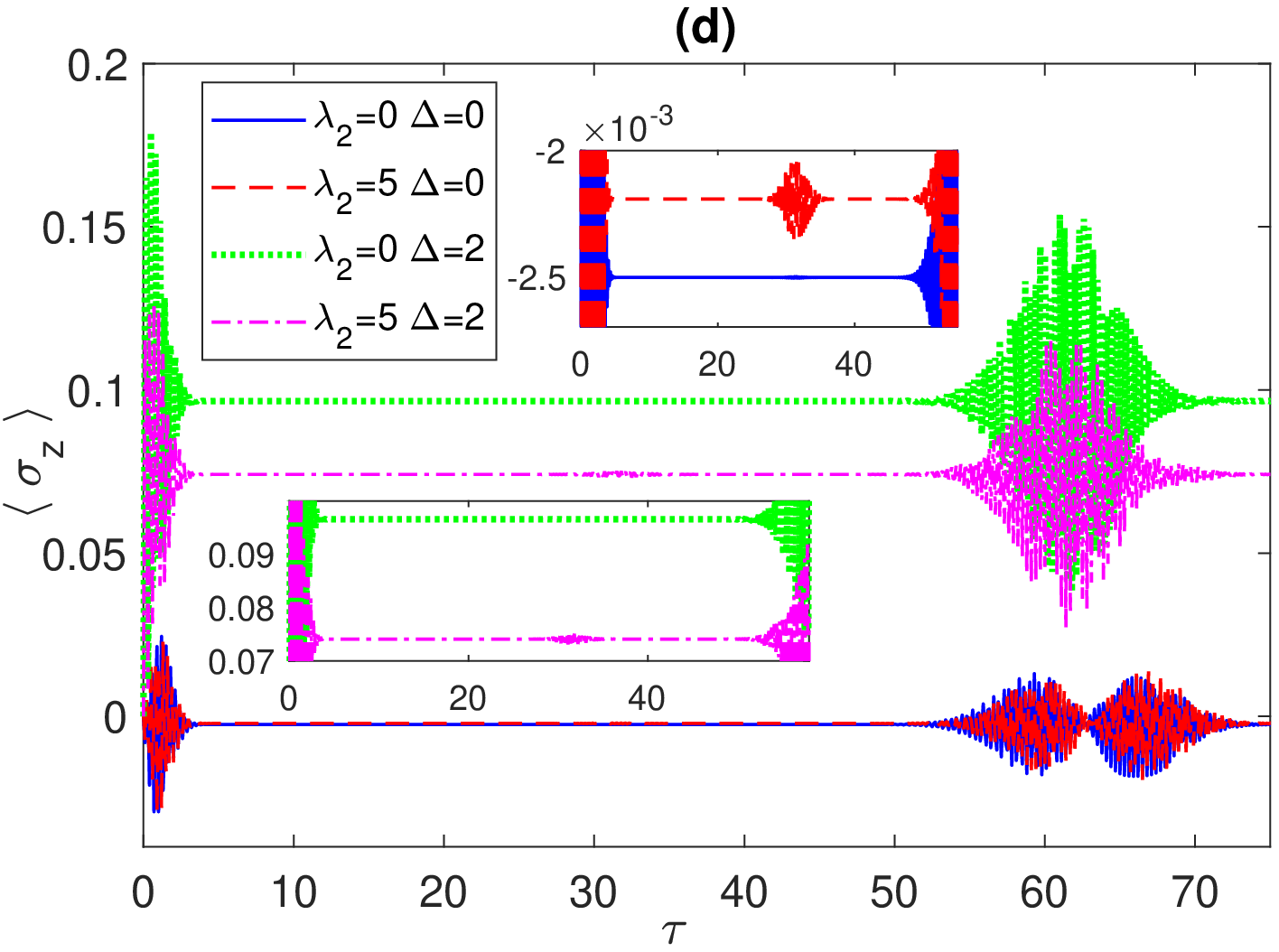}\end{subfigure}
\caption{{
Entanglement $E_f$ and population inversion $\langle \sigma_z \rangle$ versus the scaled time $\tau=\lambda_1 t$ with the two atoms are initially in a disentangled initial state  $\psi_{L}=(\vert g_{1}\rangle \vert g_{2}\rangle + \vert g_{1}\rangle \vert e_{2}\rangle + \vert e_{1}\rangle \vert g_{2}\rangle + \vert e_{1}\rangle \vert e_{2}\rangle)/\sqrt{4}$ and the field is in a coherent state:
(a) $E_f$ versus $\tau$ for various values of $\lambda_2$, $\Delta$ and the mean number of photons;
(b) $E_f$ versus $\tau$ for $\bar{n}=100$ and various values of $\lambda_2$ and $\Delta$; 
(c) $\langle \sigma_z \rangle$ versus $\tau$ for $\lambda_2=0$, $\Delta=0$ and various values of mean number of photons, and 
(d) $\langle \sigma_z \rangle$ versus $\tau$ for $\bar{n}=100$ and various values of $\lambda_2$ and $\Delta$.}}
\label{fig11}
\end{figure}

\begin{figure}[htbp]
\begin{minipage}[c]{\textwidth}
\centering   
\subfigure{\includegraphics[width=11cm]{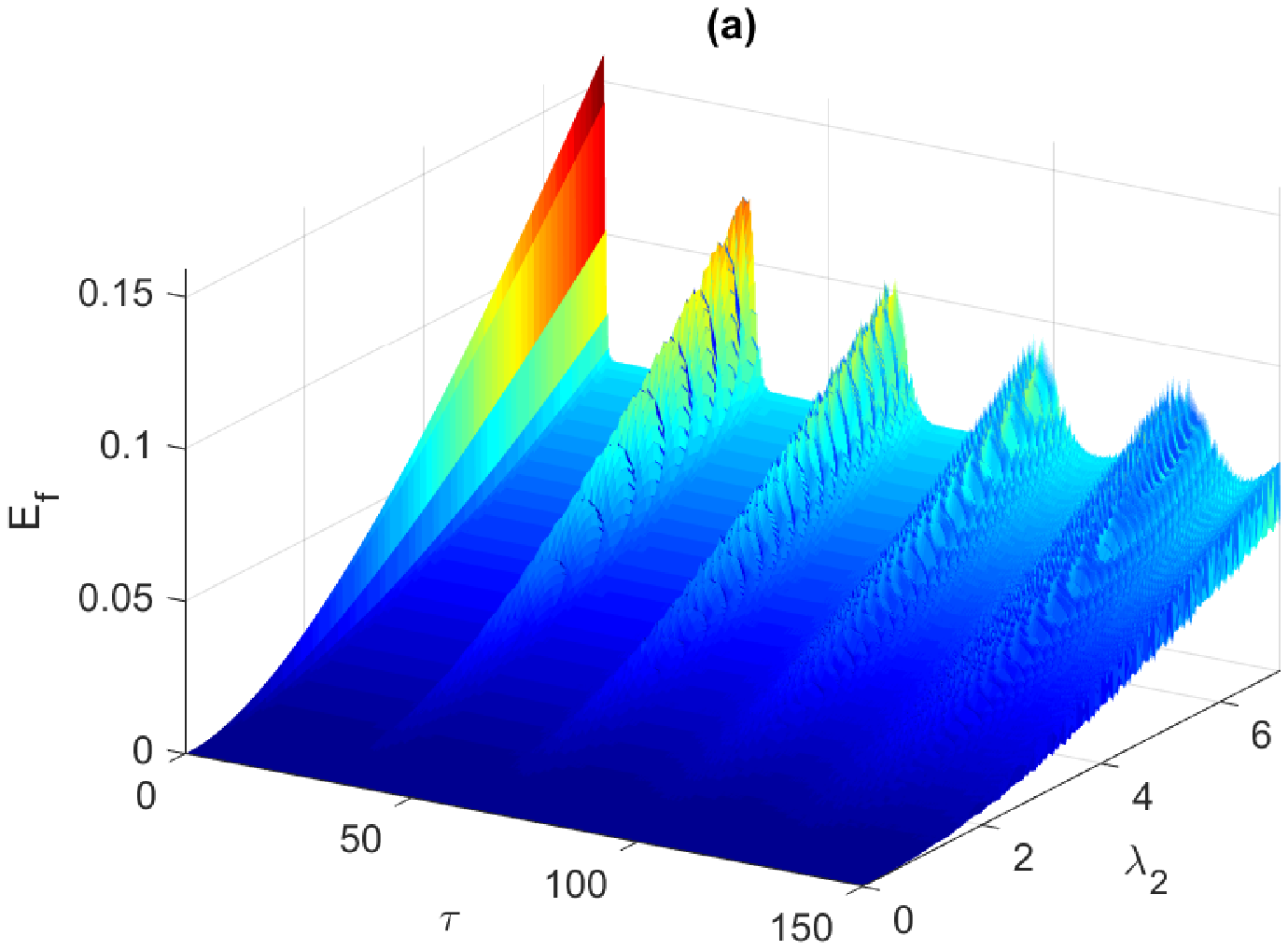}}\quad
\subfigure{\includegraphics[width=11cm]{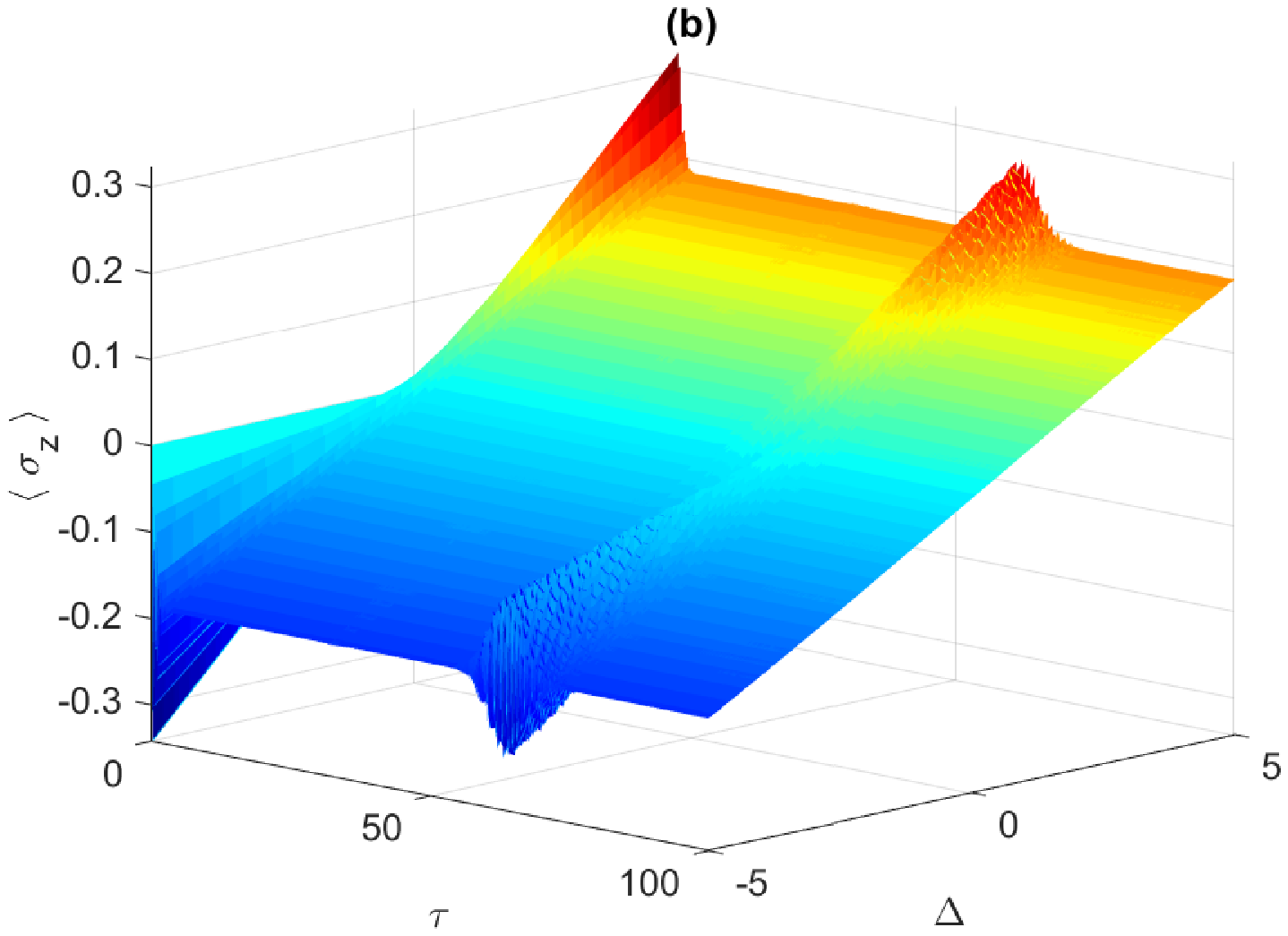}}\quad
\caption{{
(a) Entanglement versus the scaled time $\tau=\lambda_1 t$ and the coupling parameter $\lambda_2$ for $\Delta=2$, and (b) Population inversion versus the scaled time $\tau=\lambda_1 t$ and the detuning parameter $\Delta=2$ for $\lambda_2=5$, with the two atoms are initially in a disentangled initial state  $\psi_{L}=(\vert g_{1}\rangle \vert g_{2}\rangle + \vert g_{1}\rangle \vert e_{2}\rangle + \vert e_{1}\rangle \vert g_{2}\rangle + \vert e_{1}\rangle \vert e_{2}\rangle)/\sqrt{4}$ and the field is in a coherent state at $\bar{n}=100$.}}
\label{fig12}
\end{minipage}
\end{figure}

Turning on the interaction between the atoms, $\lambda_2=5$, at resonance is shown in Fig.~\ref{fig9}(c)(dashed red line). The entanglement profile is slightly different from the uncoupled case except for a big oscillation that takes place after time $\tau=60$. On the other hand, applying a non-zero detuning, $\Delta=5$, for uncoupled atoms, the entanglement oscillation shifts slightly to the right with lower peaks but the rapid oscillation appears earlier at $\tau=35$ (dotted green line). Now, turning on both detuning and coupling between the atoms, $\lambda_2=5, \Delta=5$, they enhance the entanglement and particularly raising the oscillation minima compared with the other two previous cases (dash dotted violet line). The dynamics of the population inversion starting from the initial state $\psi_e$ is illustrated in Fig.~\ref{fig9}(d). Setting the atomic coupling $\lambda_2=5$ at resonance shows a slight shift in the mean value of the population down (dashed red line) compared with the uncoupled case (solid blue line), as illustrated in the upper inset plot. Turning of the coupling and setting the detuning at $\Delta=5$, we observe a bigger shift down in the population inversion mean value and larger rapid oscillation at around $\tau=65$ (dashed green line). Now turning on both atomic coupling and non-zero detuning, $\lambda_2=5, \Delta=5$, the population mean value rises up and the rapid oscillation amplitude increases considerably at $\tau=35$ (dash dotted line), as shown in the lower inset plot. As can be noticed, the rapid oscillations of the entanglement, which takes place at its minima, are synchronized with the population revival oscillations.
The entanglement oscillation indicates that entanglement is transferred back and forth to the other subsystems, but it is not always accompanied by an atomic population revival oscillation, which means the entanglement sharing is not always mediated by atom-field energy exchange.
In Fig.~\ref{fig10}, we depict the time evolution of entanglement and population inversion over a wide range of $\lambda_2$ in (a) and (b) respectively for $\Delta=5$. One can see how increasing $\lambda_2$ spreads out and splits the population revival oscillation which also takes place at the same time for the corresponding entanglement rapid oscillation.

Finally, we consider an interesting separable initial state, which is a linear combination of all the basis states, namely  $\psi_{L}=(\vert g_{1}\rangle \vert g_{2}\rangle + \vert g_{1}\rangle \vert e_{2}\rangle + \vert e_{1}\rangle \vert g_{2}\rangle + \vert e_{1}\rangle \vert e_{2}\rangle)/\sqrt{4}$. As one can see in Fig.~\ref{fig11}(a), for uncoupled atoms at resonance with the field at $\bar{n}=20$ starting from $\psi_{L}$ (solid blue line), the entanglement of the system shows at early time a number of very narrow short spikes before completely vanishing then after a long time revives again to much shorter spikes and keeps repeating this behavior continuously. Raising the field intensity to $\bar{n}=100$ (dashed red line) doesn't lead to a noticeable change in the entanglement profile. These two cases are depicted at a magnified scale in the left inset plot of Fig.~\ref{fig11}(a). Setting a non-zero detuning in the system, $\Delta=2$, (dotted green line) enhances the entanglement value and reduces the entanglement death period. However, turning on the coupling between the two atoms, even at a very small strength $\lambda_{2}=0.2$ at zero detuning, completely eliminates the entanglement death, which shows a collapse-revival like behavior (dash-dotted violet line), where it doesn't collapse to zero value but a constant one ($\approx 7.5\times 10^{-5}$). The early time behavior of the entanglement in this two last cases is illustrated in the right inset plot of Fig.~\ref{fig11}(a) to emphasis the big impact of the atom-atom coupling on the entanglement value.
Increasing the atomic coupling further to 2 (solid blue line) then to 5 (dashed red line), the entanglement mean value increases considerably, as illustrated in Fig.~\ref{fig11}(b). Now combining atomic coupling and non-zero detuning, $\lambda_2=5, \Delta=2$ (dotted green line) the entanglement mean value decreases slightly, but higher detuning value $\Delta=5$ decreases the mean value further and makes the entanglement between the rapid oscillation periods not constant any more (dash-dotted violet line).

The effect of the radiation intensity on the atomic population is shown in Fig.~\ref{fig11}(c). One can see that applying higher intensity, where $\bar{n}=20, 50$ and finally 100 makes the collapse (constant) value approaches the zero value, also increases the collapse period and reduces the revival oscillation amplitude. Turning on the coupling, $\lambda_2=5$ at zero detuning (dash red line), the population collapse value shifts up towards the zero value and the revival oscillation amplitude increases, compared with the zero coupling case (solid blue line) as depicted in Fig.~\ref{fig11}(d) and the upper inset plot. Setting $\Delta=2$ at zero coupling, the population dynamics experiences a big shift upward above the zero value with bigger revival oscillation amplitude (dotted green line) but as we set the coupling parameter to $5$ while $\Delta=3$ the population dynamics shifts slightly down again toward the zero value (dash dotted violet line). Again, one can notice the synchronization between the entanglement and population dynamics were the constant entanglement periods correspond to the population collapse ones, while the entanglement oscillations intervals correspond to that of the revival population particularly when the atomic coupling is on. 
By looking closely at the behavior of the entanglement and the atomic population in Fig.~\ref{fig11}, one can see that the entire system starts at a separable state (the two atoms are in a completely disentangled state multiplied by the field coherent state), but the interaction between the atoms and the field, manifested as a rapid oscillation starting at $\tau=0$, triggers an entanglement rapid oscillation that eventually relaxes to a constant value that depends on the coupling strength and the detuning value. The population revival oscillation is repeated periodically and is accompanied by a rapid entanglement oscillation that doesn't lead to a new entanglement value, particularly for zero or small $\Delta$. This means in this particular case the energy exchange between the atoms and the fields do not cause entanglement transfer between the different subsystems.
In Fig.~\ref{fig12}, we show how the entanglement dynamics is very sensitive to changes in the coupling parameter $\lambda_2$, where the ESD can be completely eliminated by increasing the coupling strength, whereas the atomic population is more sensitive to variations in $\Delta$, where it changes considerably from negative to positive values as the detuning parameter is varied over a wide range from -5 to 5, as illustrated in (a) and (b) respectively.
\section{Conclusion}
We studied a system of two two-level atoms interacting with a single mode radiation field. We introduced coupling between the two atoms and considered the radiation field to be out of resonance (at non-zero detuning) with the atoms. We presented an exact analytical solution for the time evolution of the system starting from any initial state. We investigated the effect of the atom-atom coupling and the non-zero detuning separately or combined (which has not been considered before) on the atom-atom entanglement dynamics and atomic population inversion, starting from different initial states of practical interest. We showed how these two parameters can be tuned to reduce, eliminate or create entanglement sudden death (ESD) in the system, which was found to depend crucially on the initial state of the system. Particularly, we demonstrated that while one of the two interactions or both may have either negligible or weak impact on the ESD, combining them can be very effective in certain cases. Starting from an initial correlated Bell state, the time evolution of entanglement between the two uncoupled atoms at resonance with the radiation field was found to suffer sudden death (ESD) for repeated intervals with revival oscillatory peaks in between. Turning on atom-atom coupling may reduce or even eliminate the ESD  if applied at sufficient strength. Although the non-zero detuning on its own does not affect the ESD for uncoupled atoms, it does considerably when combined with the coupling of atoms and contributes significantly in the ESD removal. 
For an initial anti-correlated Bell state, the entanglement evolved to intervals of very small constant value with intermediate  revival peaks with no entanglement death observed for uncoupled atoms at resonance with the field. Nevertheless, applying a small value of detuning forces the entanglement to death intervals, but as the detuning was increased narrow intermediate peaks appeared.  On the other hand, the atom-atom coupling at zero-detuning enhanced the entanglement between the two atoms considerably. Combining the two effects had competing impacts on the bipartite entanglement, where the non-zero detuning tended to create entanglement death while the atom-atom coupling acted to remove it. Starting from a partially entangled (W-like state), the bipartite entanglement evolved to sudden death in a very similar pattern to the correlated Bell state case but with smaller ESD intervals. However, in contrary to the Bell state, the non-zero detuning reduced or even completely eliminated the ESD, as its value was increased, leading to large entanglement values, whereas the atom-atom coupling enhanced the ESD and suppressed the entanglement revival oscillation. These competing effects sustain when both interactions are present in the system at the same time.
For disentangled (separable) initial state, where both of the two atoms are in the excited state, the system never evolve to any ESD for any combination of system parameters values. Starting from that state, the entanglement showed oscillatory behavior where the non-zero detuning for uncoupled atoms raised the minima value of the entanglement oscillation and induced rapid oscillation within theses minima. Although the atom-atom coupling  at zero-detuning has a negligible effect on the entanglement pattern, it enhances the entanglement considerably when applied at non-zero detuning. Finally, we considered an interesting initial separable state, which is a linear combination of all the basis states of the system. This initial state, for uncoupled atoms at zero detuning, was found to evolve to long intervals of ESD with very small intermediate spikes. Applying a non-zero detuning has no effect on the ESD, however turning the atom-atom coupling even with a small value at zero detuning completely eliminated the ESD and lead to a collapse-revival like pattern, which did not collapse to zero but a non-zero finite value that increased as the coupling strength was increased. When a non-zero detuning was applied to coupled atoms, it reduced the constant collapse value. 
By monitoring the atomic population inversion dynamics corresponding to these different initial states, a strong synchronization was observed in each case between the population collapse-revival pattern and the entanglement dynamics. For all initial states that may evolve to ESD, for all system parameter combinations, the entanglement oscillatory revival peaks were found to be induced within the same intervals of the population revival peaks (where exchange of energy between the atoms and the radiation field takes place), whereas the sudden death (or the deviation from it) synchronized with the population collapse periods. 
In contrary, for the initial state that never evolve to any ESD under any system parameters combination, the atomic population revivals synchronized with the rapid oscillation that takes place at the minima of the entanglement oscillatory pattern, whereas the peaks of the entanglement oscillation synchronize with the population collapse periods. This means the exchange of the energy between the atoms and the fields in this case reduces the entanglement substantially.
Varying the field radiation intensity showed a big impact on the ESD time intervals at all system parameters combinations. Increasing the field intensity increased the ESD period and most of the time reduced the entanglement revival oscillation amplitude. Therefore, while all the other system parameters cannot be used to significantly modify the collapse-revival temporal pattern, the field intensity can.
The ESD behavior observed in this closed system is due to a complete entanglement transfer from the atom-atom subsystem to the atom-field subsystems in absence of any decohering effects. The system shows  entanglement revival peak after a finite time, due to an entanglement transfer back to the atom-atom subsystem. The synchronization of the revival oscillation of the atomic population with the entanglement revival peak indicates that the entanglement transfer process is mediated by the energy exchange between the atoms and the field in this case. Constrains on entanglement sharing, distribution and transfer among the different subsystems, including multipartite entanglement, of this composite system, in the presence of both of atom-atom coupling and non-zero detuning, is an interesting open question, which is currently under investigation.
\section*{Funding}
University of Sharjah, office of vice chancellor of research, grant No. 1802143060-P.
\section*{Disclosures}
The authors declare no conflicts of interest.

\end{document}